%% file: main.tex
\documentclass{article}

\usepackage[utf8]{inputenc}
\usepackage{amsmath,accents}

\usepackage[sorting=none]{biblatex}
\bibliography{bibliography}

\input{preamble.tex}
\renewcommand\Affilfont{\itshape}
\renewcommand\Authsep{, }
\renewcommand\Authand{ and }
\renewcommand\Authands{ and }
\title{
A simple yet effective ALE-FE method for the  nonlinear planar dynamics of variable-length flexible rods}
\author[1]{P. Koutsogiannakis$^1$}
\author[2]{T.K. Papathanasiou$^2$}
\author[1]{F. Dal Corso$^1$\footnote{Corresponding author: francesco.dalcorso@unitn.it}}
\affil[1]{$^1$ DICAM, University of Trento, via~Mesiano~77, Trento, Italy}
\affil[2]{$^2$ Department of Civil Engineering, Aston University, Birmingham B4 7ET, UK}

\date{\today}

\begin{document}

\input{definitions.tex}

\maketitle

\begin{abstract}
With  recent advances in variable-length structures  for use in soft actuation, energy harvesting, energy dissipation and metamaterials, the mathematical modelling and numerical simulation of physical systems with time-varying domains is becoming increasingly important. 
The planar nonlinear dynamics of one-dimensional elastic structures with  variable domain is  formulated from a Lagrangian approach by using a non-material reference frame. An Arbitrary Lagrangian-Eulerian (ALE) scheme is proposed where the domain is reparametrized based on a priori unknown configuration parameters. Based on this formulation, a Finite Element (FE) method  is developed for theoretically predicting the evolution of a rod constrained at its ends by one or two sliding-sleeves, whose position and inclination can be varied in time, and under external loadings. Finally, case studies and instability problems are investigated to assess the reliability of the proposed formulation against others available and to demonstrate its effectiveness.
With respect to previously developed methods for this type of structural problems, the present ALE-FE approach shows a strong theoretical and implementation simplicity, maintaining  an efficient and fast convergence according to the cases analyzed. An open source code realized for the present ALE-FE model is made  available for solving the nonlinear dynamics of planar systems constrained by one or two independent sliding sleeves. The present research paves the way for further extensions to easily implement solvers for the three-dimensional dynamics of flexible one- and two-dimensional structural systems with moving boundary conditions.
\end{abstract}

\noindent{\it Keywords}: Configurational forces; sliding-sleeve; nonlinear structural dynamics.

\section*{Nomenclature}
\vspace{-1mm}
\begin{minipage}[t]{0.5\textwidth}
\begin{tabular}[t]{c p{6cm}}
    ALE & Arbitrary Lagrangian-Eulerian \\
    FEM & Finite Element Method \\
    $\mathbf p$ & Vector collecting all configuration parameters \\
    $s_i$ & Configuration parameter denoting the exit of the $i$-th sliding sleeve \\
     $s$ & Arc-length spatial variable \\
     $\sigma$ & Auxiliary spatial variable \\
\end{tabular}
\end{minipage}
\begin{minipage}[t]{0.5\textwidth}
\begin{tabular}[t]{c p{6cm}}
    $\mathcal N$ & Set of all non-constrained material points \\
    $\mathcal C_i$ & Set of all material points kinematically constrained by the $i$-th sliding sleeve  \\
    $\mathcal{L}$ & Lagrangian of the structural system \\
    $\mathcal F$ & Integrant function appearing in definition of $\mathcal L$ \\
    $\mathcal B$ & Space-independent terms of $\mathcal L$ \\
\end{tabular}
\end{minipage}

\begin{minipage}[t]{0.5\textwidth}
\begin{tabular}[t]{c p{6cm}}
    $\mathcal L_{\mathcal N}$ & Terms of the Lagrangian $\mathcal L$ associated with domain $\mathcal N$ \\
    $\mathcal L_{\mathcal C_i}$ & Terms of the Lagrangian $\mathcal L$ associated with domain $\mathcal C_i$ \\
    $\mathcal E$ & Total energy of structural system \\
    $\mathcal E_{\mathcal N}$ & Energy associated to non-constrained portion of rod \\
    $\mathcal E_{C_i}$ & Energy associated to portion of rod inside $i$-th sliding sleeve\\
    $\mathcal T$ & Kinetic energy of the structural system \\
    $\mathcal T_{\mathcal N}$ & Kinetic energy of unconstrained portion of rod \\
    $\mathcal T_{\mathcal C_i}$ & Kinetic energy of rod portion inside $i$-th sliding sleeve \\
    $\mathcal V$ & Potential energy of the structural system \\
    $\mathcal V_{\mathcal N}$ & Potential energy of unconstrained portion of rod \\
    $\mathcal V_{\mathcal C_i}$ & Potential energy of rod portion inside $i$-th sliding sleeve \\
    $\mathcal W$ & Work of external forces \\
    $\boldsymbol \xi$ & Collection of fields and their space derivatives \\
    $\delta(\cdot)$ & First variation of a quantity \\
    $\dot{(\cdot)}$ & First time derivative \\
    $\ddot{(\cdot)}$ & Second time derivative \\
    $(\cdot)_{,z}$ & Derivative of quantity w.r.t $z$ \\
    $\frac{\mbox{D} (\cdot)}{\mbox{D}t}$ & Material time derivative \\
    $j$ & Jacobian of auxiliary transformation $s(\sigma)$ \\
    $\delta_{ij}$ & Kronecker delta \\
    $\mathbf{a}_i$ & Exit position of $i$-th sliding sleeve \\
    $\mathbf b_i$ & Unit vector parallel to $i$-th sliding sleeve \\
    $\mathbf n_i$ & Unit vector normal to $i$-th sliding sleeve \\
    $\theta_i$ & Angle between $i$-th sliding sleeve and $x_1$-axis \\
\end{tabular}
\end{minipage}
\begin{minipage}[t]{0.5\textwidth}
\begin{tabular}[t]{c p{7cm}}
    $\mathbf F_q$ & External forces acting on rod\\
    $\mathbf x$ & Position of the centerline of the rod \\
    $N$ & Lagrange multiplier associated with inextensibility constraint \\
    $\mathbf R_i$ & Lagrange multiplier associated with continuity of rod at $i$-th sliding sleeve exit \\
    $M_i$ & Lagrange multiplier associated with continuity of the spatial derivative of the rod centerline at $i$-th sliding sleeve exit \\
    $\mathbf C_i$ & Concentrated force at $i$-th sliding sleeve exit \\
    $\mathbf H$ & Interpolation matrix of field $\mathbf x$ \\
    $\mathbf P$ & Interpolation matrix of field $N$ \\
    $\mathcal A$ & The assembly operator of the Finite Element Method \\
    $\hat{\mathbf x}$ & Collection of degrees of freedom associated with field $\mathbf x$ \\
    $\hat{\mathbf c}$ & Collection of degrees of freedom associated with constraints and Lagrange Multipliers \\
    $(\cdot)^T$ & Transpose of vector or matrix \\
    $\beta_1, \beta_2$ & Newmark method parameters \\
    $\mathbf J$ & Jacobian matrix of discretized system of equations \\
    $N_{el}$ & Number of elements in discretization \\
    $B$ & Bending stiffness of rod \\
    $\gamma$ & Linear mass density of rod \\
    $L$ & Total length of the rod \\
    $m$ & Mass at free end of rod \\
    $\mathbf g$ & Gravity vector \\
    $\ell_0$ & Initial length of rod outside sliding sleeve \\
    $\omega$ & Angular velocity of sleeve rotation \\
    $c$ & Dissipation coefficient \\
    $\zeta$ & Non-dimensional dissipation coefficient \\
    $\mu$ & Friction coefficient \\
\end{tabular}
\end{minipage}

\section{Introduction}

At multiple scales, deployable and reconfigurable structures are the key to realising adaptive devices capable of solving engineering problems ranging from geometric constraints in transport or in working conditions and of exhibiting improved mechanical properties in response to varying loading stimuli through dramatic shape morphing. 
These types of structures can be designed by using   
origami concepts \cite{liu2022triclinic,zang2024kresling,zhang2023bistable}, programmable metamaterials \cite{chen2023programmable,gao2020shape,peng2024programming}, or highly  flexible elements \cite{mallikarachchi2014design,  pellegrinobook, pellegrinochapter}. With regard to the latter approach, the modelling of flexible structures of varying length is attracting increasing interest due to their recent connection with configurational mechanics, a theoretical framework introduced by Eshelby \cite{eshelby1956lattice,eshelby1951force} for analysing possible changes in the configuration of a solid, as for example due to a crack propagation.

Describing the deformed configuration of a variable length system requires adding one or more configuration parameters to the kinematics of the deformed part. In the case of a flexible rod partially constrained by a rigid and frictionless sliding sleeve, the configuration parameter is the relative position of the sliding sleeve exit along the rod. The sliding sleeve exit represents a \emph{moving boundary} where a discontinuity in the curvature occurs and an unexpected non-null sliding reaction force is realized, which is concentrated at the sliding-sleeve exit and has an outward direction \cite{bigoni2015eshelby}. For this particular class of structural systems, this reaction force has been independently derived by both variational and micromechanical approaches and found to be the Newtonian expression of the Eshelby force on the structural system. The  occurrence of a curvature jump at each sliding sleeve exit makes the dynamics of the rod undergoing large rotations very fast and interesting in certain instances of the phenomenon studied.

Moving boundary problems  are  of practical interest for several technological applications, as for example in passive self-tuning \cite{bukhari2021towards,miller2013experimental,staaf2018achieving}, in  oil and gas industries \cite{wicks2008horizontal,miller2015buckling}, in micro electro-mechanical systems \cite{demeio2011,goldberg2021}, soft robotics \cite{alfalahi2020concentric,alkayas2023shape,renda2021sliding},
in vibration control  \cite{dal2019nested}, in coiling \cite{jaweda2014pnas} and injection \cite{miller2015eml} processes,  and in medical catheterization \cite{duriez2006new,li2011novel}. In the last decade, research into the configurational mechanics of structures has led to the realisation of measuring devices \cite{bosi2014elastica} and of self-tunable systems \cite{koutsogiannakis2023stabilization}, as well to the establishment of novel actuation \cite{bigoni2014torsional,dal2017serpentine} and  buckling \cite{liakou2018constrained} principles. 
In addition, these advances have led to new insights into the configurational mechanics of solids for the interpretation of dislocation motion and crack advance \cite{ballarini2016newtonian}, blistering  \cite{goldberg2022material,wang2022eshelbian}, delamination \cite{venkatadri2023torsion}, penetration \cite{wen2023bending}, and ejection \cite{dal2024elastic} phenomena.

When structures with variable length are simulated with fine grained models, involving the solution of the full continuum problem and the solid-solid interaction between the rod and the sliding-sleeve, the  simulations become computationally expensive as very fine 3D meshes are needed to capture accurately the frictionless contact reaction. 
Indeed, the relative sliding between the rod and the sliding sleeve constraint implies that the contact domain is continuously evolving in time and, as a result, any computational model needs to take this into account and proper remeshing strategies have to be implemented at every time step. In classical Finite Element (FE) models the mesh is updated at the beginning of every timestep and assumed constant for the entire step. This creates a time-discrete solution having incompatibilities with the time-continuous evolution of the domain and therefore the stability and convergence of such models is not guaranteed.

\begin{figure}
    \centering
\includegraphics[width=\textwidth]{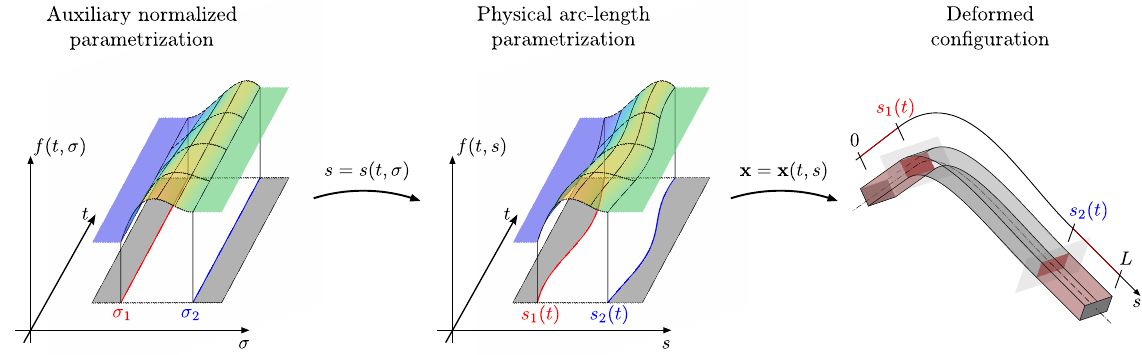}
    \caption{Auxiliary (left) and physical (right) spatial parametrizations  of a one-dimensional structure having its flexible subdomain varying in time $t$ (right). (Left) The auxiliary normalized parametrization within a time-constant domain through the space variable $\sigma\in[\sigma_1,\sigma_2]$. (Center) The physical arc-length parametrization within a time-varying domain through the space variable $s\in[s_1(t),s_2(t)]$. (Right) Deformed configuration of a one-dimensional structure of total length $L$,  fully constrained  to have null curvature within its end  subdomains $s\in\left[0,s_1(t)\right)$ and $s\in\left(s_2(t),L\right]$.}
    \label{fig:ale_spaces}
\end{figure}

A natural way for overcoming these issues is the adoption  of  a non-material description of kinematic quantities,  in analogy with the methods used in fluid mechanics \cite{donea2004arbitrary}.
Generally, an Arbitrary Lagrangian-Eulerian (ALE), also known as Mixed Eulerian-Lagrangian,  formulation can be adopted, where an underlying transformation is arbitrarily chosen, mapping the variable physical domain to a normalized auxiliary domain that remains constant in time (Fig. \ref{fig:ale_spaces}).
As a result, the mesh, defined on the normalized domain, becomes continuously variable in time when transformed to the material domain. This means that for quantities defined on material points, any time derivatives have to take into account the transformation of the domain, and equivalently that \emph{material derivatives} have to be used, similarly to the Eulerian formulation in fluid mechanics and fluid-structure interaction \cite{sanches2014fluid,wang2023arbitrary,zhang2024efficient}. 

So far, applications  have been   mainly solved within  the category of axially moving structures with prescribed boundary motion \cite{vetyukov2018non}, such as that realized in belt and pulleys mechanisms \cite{scheidl2018mixed,scheidl2021mixed,schmidrathner2022non} and in the roll forming of metal  sheets \cite{kocbay2024enhanced}.
A non-material framework for the solution of structural problems with variable-domains has been introduced in \cite{vu1995dynamics} for the simulation of the spaghetti and inverse spaghetti problems where a rod is partially inserted in one sliding sleeve and the length of the rod inside the sliding-sleeve is controlled in time, further developed in \cite{humer2013dynamic}. 
An extensive overview on non-material formulations for this problem category has been  presented in \cite{scheidl2023review}.

With regard to moving boundary problems with unprescribed motion, recently Boyer et al. presented an ALE framework for the simulation of inextensible Kirchhoff rods constrained by one sliding-sleeve using an extended Hamilton's principle \cite{boyer2022extended}, based on ideas found in \cite{behdinan1997dynamics,mciver1973hamilton}.\footnote{According to \cite{lemos2014incompleteness}, it is expected that the Hamilton-Jacobi theory cannot be properly used for the  problem under consideration since the incompleteness of this theory implies that some solution may be completely missed. Indeed,  the nonlinear problem of an elastic rod loaded at its end is analogous to that of the nonlinear pendulum, considered as an example showing  the Hamilton-Jacobi theory failing in providing all the solutions.} An ALE  method has been  also used to study the characteristic example of the dancing rod \cite{vetyukov2023dancing}, extending  the analysis of a rod's fall within a gravitational field  \cite{ARMANINI201982} by considering a distributed mass. Furthermore, it has also been used for the configurational dynamics involving
a sliding-sleeve rod system with mass concentrated at the unconstrained tip   \cite{han2022configurational}, by considering quaternions  to define the rotations of the cross-sections of the rod and dual numbers are used to obtain their final FE implementation. This methodology has been also  enhanced  in \cite{han2023configurational} to tackle the case of a rod with distributed mass interacting with one sliding-sleeve. The latter approach provides a robust ALE-FE framework for the most general case, however the dual numbers and the selection of quaternions for the kinematic parameters complicate the implementation, while the part of the rod constrained by the sleeve is simulated even though its kinematics can be directly assessed by the sliding-sleeve constraint. As a result, a lightweight ALE-FE implementation for the dynamic simulation of rods with distributed inertia and moving boundaries is not yet available.

To this purpose, inspired by the efficient dynamic FE model with fixed length developed by Bartels \cite{bartels2016simple,bartels2020numerical} enhanced by the time integration algorithm  by Papathanasiou \cite{papathanasiou2021linearised}, a FE model based on an ALE formulation is proposed for analyzing the planar dynamics of an inextensible and unshearable rod of finite length constrained at one or both ends by   frictionless sliding sleeves, whose position and inclination can be varied in time.
In this model the variable domain of the solution is defined through time-dependent configuration parameters, namely the arc-length coordinates of the exits of the sliding sleeves. These configuration parameters are a priori unknown and are modelled as degrees of freedom. The problem is formulated by splitting ab initio the rod into fully constrained and unconstrained segments, leading to the expression of the dynamics of the system solely by the fields defined on the variable domain corresponding to the unconstrained segment, and the configuration parameters. The underlying parametrization of the fields involved in the model is shown graphically in Fig. \ref{fig:ale_spaces}. For the ALE-FE  model, the weak form of the governing equations is derived from the Hamiltonian principle using a variational approach, and then the integral equations complemented by the constraints are discretized in space leading to a system of ODEs, and subsequently  the Newmark method is chosen to integrate the evolution in time. The source code of the developed software is made available \cite{Koutsogiannakis_A_simple_ALE-FEM_2024}.

The article is organized as follows. The nonlinear dynamics on variable domains is first discussed in detail together with an introduction to the ALE approach (Sect. \ref{sec:var_dom_dyn}), followed by the adaptation to the inextensible (and unshearable) sliding rod case (Sect. \ref{sec:rod_model}). Then, the derivation of the ALE-FE  model is presented in Sect. \ref{sec:discretization}. Finally, case studies and instability problems are investigated in Sect. \ref{sec:examples}, along with a comparison to  previously published results, in order to showcase the robustness of the method in cases where instabilities lead to fast dynamics.

The proposed approach provides  a powerful yet simple tool to analyze variable-domain mechanical systems for  the design of reconfigurable structures for applications in soft actuation, energy harvesting and wave mitigation. The present formulation is readily available  for an extension to treat the three-dimensional dynamics of flexible one- and two-dimensional systems with varying domains.

\section{Formulation of the nonlinear dynamics  of one-dimensional structures with movable constraints} \label{sec:var_dom_dyn}

The general background theory for solving moving boundary problems  is presented in this Section together with an introduction to  the ALE approach in Sect. \ref{repar}, adopted in  the following for the numerical resolution. The mathematical foundation for addressing the nonlinear dynamics of one-dimensional structures with  constraints movable along their domain  and the general weak form of the governing equations can be obtained through the application of the Hamiltonian
principle,
\begin{equation} \label{dalla_original}
    \int_{t_1}^{t_2} \delta \mathcal L\, \diff t = 0,\qquad
    \forall\,t_1,t_2,
\end{equation}
where $\mathcal L(t)$ is the Lagrangian of the structural system, being  $t$ the physical time. 

In the presence of movable constraints, the whole domain can be partitioned in non-overlapping time-dependent subdomains, having the constraint as boundary condition.
To this purpose, the  one-dimensional domain $[0,L]$ (being $L$ the domain size) for the space variable $s$ is reparametrized by employing a bijective (1-1 and onto) map that depends on the configuration parameter vector $\bp=\bp(t)$. The variation of integral terms is treated through the Leibniz integral rule and finally material time derivatives are introduced in order to properly perform  the integration by parts in time. This approach is analogous to Reynold's transport theorem in fluid mechanics, and yields a mathematical model for structural dynamics defined on arbitrary non-material frames of reference.

The introduced framework is exploited in Sect. \ref{sec:rod_model} to analyze structures with variable domain as the result of the presence of  sliding sleeves constraints. The latter splits the structure in two classes of variable subdomains differing in the type of imposed constraints, namely,  a variable  central subdomain $\mathcal{N}(\bp)$ with boundary conditions at its end points and  two variable end subdomains $\mathcal{C}_1(\bp)$ and $\mathcal{C}_2(\bp)$, where the kinematics of the rod is fully constrained; see Fig. \ref{fig:ale_spaces}. 
These three subdomains  are  defined  through the configuration parameters $s_1(\bp)$, and $s_2(\bp)$ describing the time-varying curvilinear coordinates along the rod where the sliding sleeves exit point is located,
\begin{equation}\label{subdomain_eqns}
    \mathcal{N}(\bp) := [ s_1(\bp), s_2(\bp) ], \quad \mathcal{C}_1(\bp) := [ 0, s_1(\bp) ), \quad \mathcal{C}_2(\bp) := ( s_2(\bp), L ].
\end{equation}

The two classes of variable subdomains requires a different treatment. In particular,  subdomain $\mathcal{N}$ is governed by two equations, one for the kinematic fields and one for the configuration parameter vector $\bp(t)$, while   subdomains $\mathcal{C}_1$ and $\mathcal{C}_2$ are respectively governed by one equation only for $\bp(t)$. This particular feature of the latter subdomains is due to the description of all kinematic fields through $\bp(t)$, meaning that the Lagrangian of the fully constrained parts of the rod can be expressed explicitly as a function of $\bp(t)$ and its time derivatives. It is noted that the dynamics of  the kinematically constrained parts ($\mathcal{C}_1$ and $\mathcal{C}_2$) is trivial in the sense that it depends only on the configurational parameter vector $\bp(t)$. Therefore, their contribution in the equations of motion can be evaluated a-priori and expressed as time-dependent concentrated loads at the moving ends, corresponding to the two sliding sleeve exits.

In the remainder of this Section, the only variation that will be evaluated is the first variation $\delta\mathcal{L}$ of the Lagrangian $\mathcal{L}$, which is the only one relevant to the dynamic problem formulation.

\subsection{First variation on non-material domains}

Following partition (\ref{subdomain_eqns}), the Lagrangian of the structural system can be evaluated as the sum of the Lagrangians  $\mathcal L_\mathcal N$, $ \mathcal L_{\mathcal C_1}$, and $\mathcal L_{\mathcal C_2}$ respectively associated with variable subdomains $\mathcal{N}(\bp)$, $\mathcal{C}_1(\bp)$, and $\mathcal{C}_2(\bp)$,
\begin{equation}
    \mathcal L = \mathcal L_\mathcal N + \mathcal L_{\mathcal C_1} + \mathcal L_{\mathcal C_2}.
\end{equation}
The Lagrangian $\mathcal L_\mathcal N$ depends on the fields  along $\mathcal N$ and terms involving evaluation of these fields at specific discrete points, while $\mathcal L_{\mathcal C_1}$ and $\mathcal L_{\mathcal C_2}$ are dependent only on $\bp(t)$,
\begin{equation}\label{functionalDependence}
    \mathcal L_\mathcal N (t, \boldsymbol{\xi}, \dot{\boldsymbol{\xi}}; \bp,\dot{\bp})  = \int_{s_1(\bp)}^{s_2(\bp)} \mathcal F (t, \boldsymbol{\xi}, \dot{\boldsymbol{\xi}}; \bp,\dot{\bp})  \,\diff s + \mathcal B_{\mathcal N}(t, \boldsymbol{\xi}_1^*,\dots, \boldsymbol{\xi}_Q^*),\qquad \mathcal L_{\mathcal C_i}( \bp, \dot\bp) = \mathcal B_{\mathcal C_i}( \bp, \dot\bp), \quad i=1,2,
\end{equation}
where  a superimposed dot represents the partial derivative in the time $t$ (defined on a \emph{material} point), $\mathcal F$ is a function of the fields involved in the problem, collected in the vector $\boldsymbol{\xi}=\boldsymbol{\xi}(t,s)$ along with their space derivatives, that defines
the system configuration and whose integration has limits that are functions of the configuration parameters $\bp$. Moreover, $\mathcal B_{\mathcal C_i}$ ($i=1,2$) are functions of $\bp$, while $\mathcal B_{\mathcal N}$ is a function of
 $\boldsymbol{\xi}_q^*=\boldsymbol{\xi}(s_q^*)$ evaluated at $Q$ specific points $s_q^*(t, \bp)\in[s_1,s_2]$, such as points of concentrated force application and of constraint location (the latter with the meaning of Lagrangian multipliers imposing kinematic boundary conditions). 
The quantities $\bp$ and $\dot\bp$ appearing after the semicolon in the argument list of the above functional $\mathcal{L}$ and function $\mathcal{F}$ represent parameters on which these are implicitly dependent only.
Next, summing up all the terms, the functional $\mathcal{L}$,
parametrized in space by the curvilinear coordinate $s$ within the time-varying compact domain $(s_1(\bp),s_2(\bp))$  can be generally defined as

\begin{equation}\label{lagrangian0}
    \mathcal{L}(t, \boldsymbol{\xi}, \dot{\boldsymbol{\xi}}; \bp,\dot{\bp}) = \int_{s_1(\bp)}^{s_2(\bp)} \mathcal{F}(t,\boldsymbol{\xi}, \dot{\boldsymbol{\xi}};\bp, \dot\bp) \diff s + \mathcal B(t,\boldsymbol{\xi}_1^*,\dots, \boldsymbol{\xi}_Q^*, \bp, \dot\bp),
\end{equation}
where $\mathcal B = \mathcal B_{\mathcal N}+\mathcal B_{\mathcal C_1}+\mathcal B_{\mathcal C_2}$.

The variation $ \delta\mathcal{L}$ of the functional $\mathcal{L}$, Eq. (\ref{lagrangian0}), can be evaluated as 
\begin{equation} \label{eq:lagrangian0}
    \delta\mathcal{L} = \delta_{\boldsymbol{\xi}}\mathcal{L}+\delta_{ \dot{\boldsymbol{\xi}}}\mathcal{L}+\delta_{\bp}\mathcal{L}+\delta_{\dot\bp}\mathcal{L}, 
\end{equation}
where $\delta_{\boldsymbol{\xi}}\mathcal{L}$, $\delta_{ \dot{\boldsymbol{\xi}}}\mathcal{L}$, $\delta_{\bp}\mathcal{L},$ and $\delta_{\dot\bp}\mathcal{L}$ are variations of the functional $\mathcal{L}$ with respect to the  single corresponding variable, namely
\begin{equation} \label{eq:lagrangian0single}
    \delta_{\bz}(\cdot)=(\cdot)_{,\bz}\cdot\delta \bz, 
\end{equation}
where  the notation $(\cdot)_{,\bz}$ is introduced to denote the partial derivative of a quantity with respect to a variable $\bz$; i.e. $(\cdot)_{,\bz} = \partial (\cdot) / \partial \bz$.

Considering the introduced assumptions about the functionals, Eq. (\ref{functionalDependence}), the following derivatives vanish
\begin{equation}
    \mathcal F_{,\bp}= \mathcal F_{,\dot\bp} =\mathcal B_{\mathcal N,\bp} = \mathcal B_{\mathcal N,\dot\bp} = \mathbf{0}, \qquad
    \mathcal B_{\mathcal C_i,\boldsymbol{\xi}^*_q} = \mathbf{0}, \quad i=1,2,\qquad q=1,...,Q,
    \end{equation}
and, by using  the Leibniz integral rule and the chain rule, the variations $\delta_{\boldsymbol{\xi}}\mathcal{L}$, $\delta_{ \dot{\boldsymbol{\xi}}}\mathcal{L}$, $\delta_{\bp}\mathcal{L}$, and $\delta_{\dot\bp}\mathcal{L}$ reduce to
\begin{equation}
    \begin{array}{lll}
       \displaystyle \delta_{\boldsymbol{\xi}}\mathcal{L}= \int_{s_1(\bp)}^{s_2(\bp)} \mathcal{F}_{,\boldsymbol{\xi}} \cdot \delta \boldsymbol{\xi} \diff s+\sum_{q=1}^Q \mathcal{B}_{,\boldsymbol{\xi}^*_q}\cdot 
       \delta \boldsymbol{\xi}^*_q, \qquad
       & \displaystyle 
        \delta_{ \dot{\boldsymbol{\xi}}}\mathcal{L}=
        \int_{s_1(\bp)}^{s_2(\bp)} \mathcal{F}_{, \dot{\boldsymbol{\xi}}} \cdot \delta  \dot{\boldsymbol{\xi}} \diff s,
        \\
        \displaystyle \delta_{\bp}\mathcal{L}=\left\{\bigg[\mathcal{F}\ s_{,\bp} \bigg]_{s_1}^{s_2}+ \mathcal  B_{,\bp} + \sum_{q=1}^Q\mathcal{B}_{,\boldsymbol{\xi}_q^*}\cdot \boldsymbol{\xi}^*_{q,s_q^*}\ s^*_{q,\bp} \right\}\cdot \delta\bp,
        \qquad\qquad
       & \delta_{\dot\bp}\mathcal{L}=\mathcal  B_{,\dot\bp} \cdot \delta\dot\bp,
    \end{array}
\end{equation}
so that the first variation $ \delta\mathcal{L}$ results to
\begin{equation} \label{eq:lagrangian}
    \delta\mathcal{L} = \underbrace{\int_{s_1(\bp)}^{s_2(\bp)} \left(\mathcal{F}_{,\boldsymbol{\xi}} \cdot \delta \boldsymbol{\xi} + \mathcal{F}_{,\dot{\boldsymbol{\xi}}} \cdot \delta \dot{\boldsymbol{\xi}}\right) \diff s +\sum_{q=1}^Q \mathcal{B}_{,\boldsymbol{\xi}^*_q} \cdot \delta \boldsymbol{\xi}^*_q }_{\mbox{\footnotesize{\lq fixed domain' terms}}}+ \underbrace{\left\{\bigg[\mathcal{F}\ s_{,\bp} \bigg]_{s_1}^{s_2}+ \mathcal  B_{,\bp} + \sum_{q=1}^Q\mathcal{B}_{,\boldsymbol{\xi}_q^*}\cdot \boldsymbol{\xi}^*_{q,s_q^*}\ s^*_{q,\bp} \right\}\cdot \delta\bp + \mathcal  B_{,\dot\bp} \cdot \delta\dot\bp}_{\mbox{\footnotesize{configuration parameter-dependent terms}}}.
\end{equation}

It is finally noted that the following identities relevant to  the evaluation of the derivatives of $\delta\mathcal B$ hold
\begin{equation}
    \mathcal B_{,\boldsymbol{\xi}_q^*} = \mathcal B_{\mathcal N,\boldsymbol{\xi}_q^*}, \qquad
    \mathcal B_{,\bp} = \mathcal B_{\mathcal C_1,\bp}  + \mathcal B_{\mathcal C_2,\bp}, \qquad
    \mathcal B_{,\dot\bp} = \mathcal B_{\mathcal C_1,\dot\bp}+ \mathcal B_{\mathcal C_2,\dot\bp},
\end{equation}
which can be obtained by considering the expressions in Eq.  (\ref{functionalDependence}) for $\mathcal B_{\mathcal N}$, $\mathcal B_{\mathcal C_1}$, and $\mathcal B_{\mathcal C_2}$.

\subsection{Reparametrization to a time-independent \lq auxiliary' domain}
\label{repar}
The first variation $\delta \mathcal{L}$, Eq. \eqref{eq:lagrangian}, involved in the Hamiltonian principle \eqref{dalla_original} contains time derivative variations $\delta \dot{\boldsymbol{\xi}}$ and $\delta\dot{\bp}$, which need to be treated for obtaining the weak form of the governing equations. Although standard for time-independent domains, their integration by parts for time-dependent domains requires special attention.

More specifically,  following the ALE approach, the spatial integral has to be transformed to a time-independent \lq auxiliary' domain, to allow the change in the integration order between  time and space. For this reason, a diffeomorphic map is defined such that a non-varying compact domain for the \lq auxiliary' spatial variable $\sigma$ is mapped into a varying domain for the \lq physical'  spatial variable $s$,

and without loss of generality and for simplicity, the following linear relation is adopted
\begin{equation}\label{mappanos}
    s(t, \sigma) = s_1(t) + (s_2(t)-s_1(t))\ \sigma,
\end{equation}
which maps $\sigma\in[0,1]$ onto $s\in [s_1(t), s_2(t)]$. As a consequence, after integration by parts of the terms containing time-derivatives of variations, the Hamiltonian principle (\ref{dalla_original}) for variable domains can be rewritten as
\begin{equation} \label{eq:var_princ}
\begin{multlined}
    \left[\int_{s_1}^{s_2} \mathcal{F}_{,\dot{\boldsymbol{\xi}}}\!\cdot \delta \boldsymbol{\xi} \diff s \right]_{t_1}^{t_2} - \int_{t_1}^{t_2} \int_{s_1}^{s_2} \left\{ \left[\frac{\mbox{D}\ }{\mbox{D} t}\mathcal{F}_{,\dot{\boldsymbol{\xi}}} + j^\prime\ \mathcal{F}_{,\dot{\boldsymbol{\xi}}} -\mathcal{F}_{,\boldsymbol{\xi}} \right] \!\cdot  \delta \boldsymbol{\xi} + \dot s\ \mathcal{F}_{,\dot{\boldsymbol{\xi}}} \!\cdot \ \delta \boldsymbol{\xi}_{,s} \right\} \diff s \diff t  + \int_{t_1}^{t_2} \bigg[\mathcal{F}\ s_{,\bp} \bigg]_{s_1}^{s_2} \cdot \delta\bp \diff t \\
      + \bigg[ \mathcal B_{,\dot\bp} \cdot \delta\bp \bigg]_{t_1}^{t_2} + \int_{t_1}^{t_2} \bigg\{ \sum_{q=1}^Q \mathcal{B}_{,\boldsymbol{\xi}^*_q} \cdot \delta \boldsymbol{\xi}^*_q + \bigg[ \sum_{q=1}^Q \mathcal{B}_{,\boldsymbol{\xi}^*_q}\cdot \boldsymbol{\xi}_{q,s}^*\ s^*_{q,\bp} + \mathcal  B_{,\bp} - (\mathcal B_{,\dot\bp})_{,t}\bigg] \cdot \delta\bp \bigg\} \diff t = 0,
\end{multlined}
\end{equation}
where $j(t)$ and $\mbox{D}(\cdot)/\mbox{D} t$ respectively denote the transformation Jacobian  and the \emph{material} time derivative operator, \begin{equation}
    j(t) = s_{,\sigma}(s), \quad \frac{\mbox{D} (\cdot)}{\mbox{D} t} = (\cdot)_{,t} + (\cdot)_{,s} \ \dot s,
\end{equation}
while $j^{-1}(t)$ is the inverse of the transformation Jacobian and $j^\prime$ denotes the gradient of the time derivative of the transformation Jacobian 
\begin{equation}
    j^{-1}(t)=\sigma_{,s}(t),\qquad
    j^\prime = \left(j\right)^{\cdot}\cdot j^{-1},
\end{equation}
which, by considering the adopted linear mapping (\ref{mappanos}), reduces to  
\begin{equation}
    j^{-1}(t) = \dfrac{1}{j(t)},
    \qquad
    j^\prime =  (s_{,t})_{,s}.
\end{equation}

Since the Hamiltonian principle (\ref{eq:var_princ}) holds for every admissible variations $\delta \boldsymbol{\xi}$, and $\delta \bp$, 
that satisfies
\begin{equation} \label{adm_var}
    \delta \boldsymbol{\xi}(t)=\delta \bp(t)=\mathbf{0},\qquad \mbox{for}\,\, t=t_1,t_2,
\end{equation}
the  weak form of the governing equations follows as the following differential system
\begin{equation} \label{eq:variation_system}
    \left\{
    \begin{aligned}
        & \begin{multlined}
             \int_{s_1}^{s_2} \left[ \left(\frac{\mbox{D}\ }{\mbox{D} t}\mathcal{F}_{,\dot{\boldsymbol{\xi}}} + j^\prime\ \mathcal{F}_{,\dot{\boldsymbol{\xi}}} - \mathcal{F}_{,\boldsymbol{\xi}} \right)\! \cdot\delta \boldsymbol{\xi} + \dot s\ \mathcal{F}_{,\dot{\boldsymbol{\xi}}} \!\cdot \ \delta \boldsymbol{\xi}_{,s} \right] \diff s 
            - \sum_{q=1}^Q \mathcal{B}_{,\boldsymbol{\xi}^*_q} \cdot \delta \boldsymbol{\xi}^*_q = 0,
        \end{multlined} \\[0.7em]
        &\left[
        \left. \bigg(\mathcal{F}\ s_{,\bp} \bigg)\right|_{s_1}^{s_2} + \sum_{q=1}^Q \mathcal{B}_{,\boldsymbol{\xi}^*_q}\cdot \boldsymbol{\xi}_{q,s}^*\ s^*_{q,\bp} + \mathcal  B_{,\bp} - (\mathcal B_{,\dot\bp})_{,t} \right] \cdot \delta\bp = 0.
    \end{aligned}
    \right.
\end{equation}
It can be noted that the governing equations (\ref{eq:variation_system}) represent a system of one partial differential equation
and one ordinary differential equation, where the latter expresses  interface conditions at the moving boundaries $s_1$ and $s_2$.
It is finally noted that the equations of motion can also be independently obtained through dynamic equilibrium by considering the presence of a configurational force expressed by an outward nonlinear tangential reaction force at each frictionless sliding sleeve exit.

\section{Structural dynamics of inextensible geometrically nonlinear rods with ends constrained by sliding sleeves} \label{sec:rod_model}

The planar nonlinear dynamics of a rod constrained at its two edges by sliding sleeves is introduced by using the framework established in Sect. \ref{sec:var_dom_dyn}. The rod, of uniform bending stiffness $B$ and linear mass density $\gamma$, is modelled as inextensible, meaning that it displays null axial strain, and unshearable, meaning that the cross sections remain orthogonal to the rod's axis. The  straight configuration is considered as the unloaded state for the rod. The cross section along the rod is identified through the curvilinear coordinate $s\in[0,L]$, being $L$ the total length.

The case of rods  constrained  by sliding sleeves at both ends is analyzed first, while the case of only one side constrained by a sliding sleeve and the other free is then obtained as a simplification. 

\subsection{Rod constrained at each of its edges by two independent sliding sleeves} \label{sec:two_sleeve}

\begin{figure}
    \centering
    \includegraphics[width=102mm]{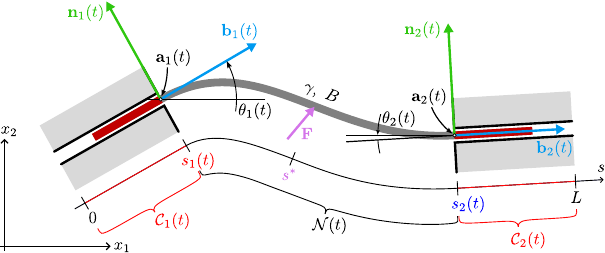}
    \caption{Sketch of the planar structural system with an unconstrained  time-varying domain $\mathcal{N}(t)$. The  one-dimensional structure is constrained at its ends by two independent sliding sleeves controlled in the time $t$ through the position vector $\bp_i(t)$  and the unit tangent $\mathbf{b}_i(t)$, the latter inclined by the angle $\theta_i(t)$ (positive when counter-clockwise) with respect to the $x_1$ axis ($i=1,2$). The rod has linear mass density $\gamma$, total length $L$ and is flexible under a bending stiffness $B$, but inextensible and unshearable. The position of the sliding sleeves exit along the rod is measured through the curvilinear coordinate $s_1(t)$ and $s_2(t)$, whose in turn define the fully-constrained time-varying  subdomains $\mathcal{C}_1(t)$ and $\mathcal{C}_2(t)$.}
    \label{fig:two_sleeve_drawing}
\end{figure}

Following Sect. \ref{sec:var_dom_dyn}, the curvilinear coordinate domain, $s\in[0,L]$, can be  split in three variable subdomains, two associated with the rod's portions fully constrained by the sliding sleeves (see Fig. \ref{fig:two_sleeve_drawing}), named $\mathcal{C}_1(t)$ and $\mathcal{C}_2(t)$, and one associated with the unconstrained rod's portion, $\mathcal{N}(t)$, Eq. (\ref{subdomain_eqns}) with $s_1(t)$ and $s_2(t)$ as configuration parameters,
\begin{equation}
    \mathcal C_1(t) = [0, s_1(t)],\quad \mathcal C_2(t) = [s_2(t),L],\quad \mathcal N(t) = [s_1(t), s_2(t)].
\end{equation}

The presence of a (constant) gravitational distributed  force $\gamma \bg$ along the entire rod and of concentrated external forces $\bF_{q}(t)$ at $Q$ curvilinear coordinates $s^*_q(t,\bp)$  is considered ($q=1,..,Q$). 

The position  in time $t$ of a material point  along the rod, associated with   the curvilinear coordinate $s$, is described by the vector $\bx(t,s)$ collecting the two coordinates in the Cartesian $x_1$--$x_2$ reference system. The inextensibility constraint dictates that the spatial derivative of $\bx$, corresponding to a tangent vector to the rod axis, has a unit modulus at every material point, namely
\begin{equation} \label{eq:inext}
    ||\bx_{,s}(t,s)|| = 1,\qquad \forall s\in[0,L],
\end{equation}
where $||\cdot||$ represents the Euclidean norm.

Moreover, the  position $\bx(t,s)$ of material points of the rod inside the two sliding sleeves is constrained by (repeated index is not summed) \begin{equation} 
\begin{aligned}\label{positionconstrained}
    & \bx(t,s) = \ba_i(t) + \left[s-s_i(t)\right]\ \bb_i(t),\quad s\in \mathcal{C}_i(t),\quad i=1,2,
\end{aligned}
\end{equation}
where $\ba_1(t)$ and $\ba_2(t)$ are the positions of the exit of the two sliding sleeves, while  $\bb_1(t)$, and $\bb_2(t)$ are the unit vectors parallel to the sliding-sleeves, according to the direction of the rod's tangent, inclined at angles $\theta_1$ and $\theta_2$ (positive angle when counterclockwise) with respect to the $x_1$ axis , Fig. \ref{fig:two_sleeve_drawing}, namely
\begin{equation}
    \mathbf{b}_i(t)=\left(
    \begin{array}{cc}
         \cos\theta_i(t)\\
         \sin\theta_i(t)
    \end{array}\right).
\end{equation}
From the constrained position fields (\ref{positionconstrained}), the fields of material velocity $\dot\bx(t,s)$ and acceleration $\ddot\bx(t,s)$ are given in the fully constrained subdomains $\mathcal{C}_1$ and $\mathcal{C}_2$ by
\begin{equation}\label{velocity_constrained}
    \begin{aligned}
        & \dot\bx(t,s) = \dot\ba_{i}(t) - \dot s_{i}(t)\ \bb_i(t) + \left[s-s_i(t)\right]\ \dot\bb_{i}(t), \\
        & \ddot\bx(t,s) = \ddot\ba_{i}(t) - \ddot s_{i}(t)\ \bb_i(t) - 2\, \dot s_{i}(t)\ \dot\bb_{i}(t) + \left[s-s_i(t)\right]\ \ddot\bb_{i}(t),
    \end{aligned}
    \qquad s\in \mathcal{C}_i(t),\quad i=1,2.
\end{equation}
From Eq. (\ref{positionconstrained}), the continuity along the rod of  both the position $\bx(t,s)$ and the tangent $\bx_{,s}$ leads  the two following additional constraints at each sliding sleeve
\begin{equation}
    \bx\left(t,s_i(t)\right) - \ba_i(t) = \b0, \quad \bx_{,s}(t,s_i(t))\cdot \bn_i(t) = 0,\quad i=1,2,
\end{equation}
where the unit vector $\bn_i(t)$ is the normal to the $i$-th sliding sleeve, namely
\begin{equation}
    \mathbf{n}_i(t)=\left(
    \begin{array}{cc}
       -\sin\theta_i(t)\\  \cos\theta_i(t)
    \end{array}\right).
\end{equation}

The kinetic energy $\mathcal T_{Ci}$ and the total potential energy $\mathcal V_{Ci}$ associated to the  rod portion inside the $i$-th sliding sleeve (subdomain $\mathcal{C}_i$) are given by
\begin{equation}
    \mathcal T_{Ci} = \frac{\gamma}{2}\int_{\mathcal{C}_i(t)}  \dot\bx^2(t,s)\, \diff s,\quad 
     \mathcal V_{Ci} = -\gamma\, \bg \cdot \int_{\mathcal{C}_i(t)}  \bx(t,s) \, \diff s,
     \quad i=1,2.
\end{equation}

The kinetic $\mathcal T_{Ci}\left(s_i(t),\dot s_i(t)\right)$ and potential energy $\mathcal V_{Ci}\left(s_i(t)\right)$ ($i=1,2$) can be evaluated by  substituting the expressions for the rod position $\bx(t,s)$ and velocity $\dot\bx(t,s)$, Eqs.  (\ref{positionconstrained}) and (\ref{velocity_constrained})$_1$, as functions only of  the configuration parameters $s_1(t)$ and $s_2(t)$ and their velocity $\dot{s}_1(t)$ and $\dot{s}_2(t)$ as follows ($\delta_{ij}$ is the Kronecker delta, repeated index is not summed)
\begin{equation}
\begin{array}{lll}
    \mathcal{T}_{Ci}(t) =& \displaystyle\frac{\gamma}{2}\left\{ \left[ \left( \dot\ba_{i} - \dot s_{i} \bb_i \right)^2 - 2\, s_i\, \dot\ba_{i}\cdot \dot\bb_{i} + 2\,s_i\, \dot s_{i}\, \bb_i\cdot \dot\bb_{i}  + s_{i}^2\, \dot\bb_{i}^2  \right] \left(\delta_{2i} L-(-1)^{i}\, s_i\right) \right. \\
        &\displaystyle\left.+ \left[ \dot\ba_{i}\cdot \dot\bb_{i} - \dot s_{i}\, \bb_i\cdot \dot\bb_{i}  - s_i\, \dot\bb_{i}^2  \right] \left(\delta_{2i} L^2-(-1)^{i}\, s_i^2\right) + \frac{\dot\bb_{i}^2}{3} \left(\delta_{2i} L^3-(-1)^{i}\, s_i^3 \right) \right\},\\
     \mathcal V_{Ci}(t) =& \displaystyle
    -\gamma \bg\cdot\left[ (\ba_i - s_i\, \bb_i) \left(\delta_{2i} L-(-1)^{i}\, s_i\right) + \frac{\bb_i}{2} \left(\delta_{2i} L^2-(-1)^{i}\, s_i^2\right) \right],
\end{array}\qquad
i=1,2.
\end{equation}

The kinetic  $\mathcal T_\mathcal{N}\left(s_1(t),s_2(t), \dot\bx(t,s) \right)$ 
and the potential $\mathcal V_\mathcal{N}\left(s_1(t),s_2(t), \bx(t,s) \right)$ energies of the central portion of the rod  $\mathcal{N}$ are given by 
\begin{equation}
    \mathcal T_\mathcal{N} =  \frac{\gamma}{2}\int_{s_1(t)}^{s_2(t)}\dot\bx^2(t,s) \, \diff s,
    \qquad 
    \mathcal V_\mathcal{N} =  \frac{B}{2}\int_{s_1(t)}^{s_2(t)} \bx_{,ss}^2(t,s) \diff s
    - \gamma\, \bg\cdot\int_{s_1(t)}^{s_2(t)} \bx(t,s) \diff s
    - \sum\limits_q^Q \bF_q(t)\cdot \bx(t,s_q^*).
\end{equation}

The Lagrangian $\mathcal L$ of the whole system is provided by 
\begin{equation}\label{LagrangianSystemTOT}
\begin{multlined}
    \mathcal L = \mathcal T_\mathcal{N} + \mathcal T_{\mathcal C_1} + \mathcal T_{\mathcal C_2} - \mathcal V_\mathcal{N} - \mathcal V_{\mathcal C_1} - \mathcal V_{\mathcal C_2} \\ \qquad - \int_{s_1}^{s_2} \dfrac{N}{2} \left( \bx_{,s}^2 - 1 \right) \diff s - \bR_1\cdot\left[ \bx(s_1) - \ba_1 \right] - M_1\,\bn_1 \cdot \bx_{,s}(s_1) - \bR_2\cdot\left[ \bx(s_2) - \ba_2 \right] - M_2\,\bn_2 \cdot \bx_{,s}(s_2),
\end{multlined}
\end{equation}
where $N(t,s)$ is a Lagrange multiplier field associated with the inextensibility constraint defined on $(s_1(t),$ $ s_2(t))$, and $\bR_1(t)$, $\bR_2(t)$, $M_1(t)$, and $M_2(t)$ are Lagrange multipliers associated with the continuity of the rod at the sliding sleeve exits. These Lagrangian multipliers have the physical meaning of  internal axial force ($N(t,s)$) along the central portion of the rod and of  the balancing reaction force vector ($\bR_i(t)$) and  moment ($M_i(t)$) at the $i$-th sliding sleeve exit respectively. It is noted that, as shown at the end of this SubSection,  the force $\mathbf{R}_i$ has to not be confused with the concentrated reaction $\mathbf{C}_i$ realized at the $i$-th sliding sleeve exit.

Finally, given the Lagrangian  $\mathcal{L}$ \eqref{LagrangianSystemTOT},  the following weak form of the governing equations of the structural system can be obtained from Eq. \eqref{eq:variation_system} as
\begin{equation} \label{eq:gov_eq}
    \begin{aligned}
        & \begin{multlined}
            \int_{s_1(t)}^{s_2(t)} \left\{ \gamma \left[ \frac{\mbox{D}^2 \bx }{\mbox{D} t^2} - \ddot s\ \bx_{,s} + j^\prime\  \frac{\mbox{D} \bx }{\mbox{D} t} - \dot s \left( \frac{\mbox D \bx}{\mbox D t} \right)_{\!\!,s} - \bg \right]\cdot \delta\bx \right.
            \\ \qquad\qquad
            + \left[ \gamma\, \dot s\, \left( \frac{\mbox{D} \bx }{\mbox{D} t} - \dot s\ \bx_{,s} \right) + N\,\bx_{,s} \right]\cdot \delta \bx_{,s}
            + B\,\bx_{,ss}\cdot\delta\bx_{,ss} \bigg\} \diff s 
            \\ \qquad\qquad\qquad \qquad\qquad\qquad
            -\sum\limits_q^Q \bF_q(t)\cdot \delta\bx(t,s_q^*) + \sum\limits_{i=1}^2 \left[ \bR_i \cdot \delta\bx(s_i) + M_i\,\bn_i\cdot\delta\bx_{,s}(s_i) \right] = 0,
        \end{multlined} \\
        & \begin{multlined}
            \left\{ \left( \frac{\gamma}{2} \dot\bx^2(s_i) - \frac{B}{2}\bx_{,ss}^2(s_i) + \gamma \bg\cdot \bx(s_i) \right) (-1)^i + \mathcal T_{Ci,s_i} - (\mathcal{T}_{Ci,\dot s_{i}})_{,t} - \mathcal V_{Ci,s_i} + (\mathcal{V}_{Ci,\dot s_{i}})_{,t}\right. \\
            \qquad\qquad\qquad\qquad\qquad\qquad\qquad\qquad\qquad\qquad\qquad
            - \bR_i \cdot \bx_{,s}(s_i) - M_i\,\bn_i\cdot\bx_{,ss}(s_i) \bigg\} \delta s_i = 0 , \quad
            i=1,2,
        \end{multlined}
    \end{aligned}
\end{equation}
where the identity
\begin{equation}
    \frac{\mbox{D}\dot \bx }{\mbox{D} t}  = \frac{\mbox{D}^2 \bx }{\mbox{D} t^2} - \ddot s\, \bx_{,s} - \dot s \left(\frac{D\bx}{Dt}\right)_{,s} + \dot s\ j^\prime\, \bx_{,s},
\end{equation}
has been used in order to express the weak form using quantities defined on the moving mesh and the function values at $s_i$ should be interpreted as the evaluation at $s_1^+$ for $i=1$ (at $s_2^-$ for $i=2$), providing the function value at the coordinate $s_1$ ($s_2$) approaching from the right (left).
The weak form \eqref{eq:gov_eq}  of the governing equations is complemented by the following algebraic constraints 
\begin{equation}
    \begin{array}{llll}
        & \ds \int_{s_1(t)}^{s_2(t)} \frac{1}{2}\ \delta N\ (\bx_{,s}\cdot\bx_{,s}-1)\, \diff s = 0, \quad
        & \delta\bR_1 \cdot \left(\bx(t,s_1) - \ba_1(t) \right) = 0, \qquad
        & \delta\bR_2 \cdot \left(\bx(t,s_2) - \ba_2(t) \right) = 0,      \\[0.5em]
        & \delta M_1 \ \bn_1(t)\cdot \bx_{,s}(t,s_1) = 0, \qquad
        & \delta M_2 \ \bn_2(t)\cdot \bx_{,s}(t,s_2) = 0.  \qquad
        &
    \end{array}
\end{equation}

It is noted here that in addition to the admissibility conditions \eqref{adm_var} the variations $\delta\bx$ need to satisfy also the inextensibility condition
\begin{equation}
    \delta\bx_{,s}(t,s)\cdot \bx_{s}(t,s) = 0,\quad \forall t, \quad\forall s\in\mathcal N(t).
\end{equation}

\paragraph{Strong form.}
Assuming $B$ constant, sufficient regularity for $\bx$ and $N$, and using integration by parts for the terms of Eq. \eqref{eq:gov_eq}$_1$ involving space derivatives of the variations $\delta\bx$, the problem can be strongly formulated in the non-material frame of reference as
\begin{equation}
    \gamma \left[ \frac{\mbox D \dot\bx}{\mbox D t} - \dot s \cdot \left( \frac{\mbox D \bx}{\mbox D t} \right)_{\!\!,s} \right] + \left(\gamma \dot s \dot s_{,s} - N_{,s} \right) \bx_{,s} + \left( \gamma \dot s^2 - N \right) \bx_{,ss} + B\, \bx_{,ssss}  - \gamma\,\bg = 0, \quad \forall\ s=s(t,\sigma),\ t,
\end{equation}
or, equivalently, in the material framework as
\begin{equation}\label{panosisstrong}
     B\, \bx_{,ssss}- \left( N\, \bx_{,s} \right)_{,s}  =\gamma \,\left( \bg -\ddot\bx \right) , \quad \forall\ s,\ t,
\end{equation}
complemented by the natural boundary conditions
\begin{equation}
\begin{array}{lll}\label{panoshasboundaries}
    & \bR_1 - \left[ \gamma\,\dot s \,\dot\bx + N\, \bx_{,s} - B\,\bx_{,sss} \right]_{s_1^+} = \b0, \qquad \qquad
    & M_1\,\bn_1 - B\, \bx_{,ss}(s_1^+) = \b0, \\[1em]
    & \bR_2 + \left[ \gamma\,\dot s  \,\dot\bx + N\, \bx_{,s} - B\,\bx_{,sss} \right]_{s_2^-} = \b0, \qquad \qquad
    & M_2\,\bn_2 - B\, \bx_{,ss}(s_2^-) = \b0,
\end{array}
\end{equation}
in addition to Eq. \eqref{eq:gov_eq}$_2$
and under the inextensibility constraint, Eq. \eqref{eq:inext}. 

Eqs. \eqref{eq:gov_eq}$_2$ and \eqref{panoshasboundaries} allow to write the transverse and sliding component of the forces $\bR_i$ ($i=1,2$) as
\begin{equation} \label{eq:configforce}
    \begin{aligned}
        & \bR_i \cdot \bn_i = (-1)^i B\, \bx_{,sss} (s_i)\cdot \bn_i, \quad
        & \bR_i \cdot \bb_i = (-1)^i \frac{M_i^2}{2\,B} + T_i,
    \end{aligned}
\end{equation}
where $ T_i$ models the sliding forces developed due to the motion of the rod's portion fully constrained by the $i$-th sliding sleeve and is a function of the vectors $\ba_i$, and $\bb_i$ ($i=1,2$)
\begin{equation}
    T_i = T_i(\ba_i, \dot\ba_i, \ddot\ba_i, \bb_i, \dot\bb_i, \ddot\bb_i) = \mathcal T_{Ci,s_i} - (\mathcal{T}_{Ci,\dot s_{i}})_{,t} - \mathcal V_{Ci,s_i} + (-1)^i \gamma \left( \dfrac{\dot s_1^2}{2} + \bg\cdot\ba_i \right).    
\end{equation}
 Therefore, $\bR_i$ balances the sum of the concentrated force $\mathbf{C}_i$ at the $i$-th sliding sleeve exit, as obtained in \cite{ARMANINI201982}, and the additional sliding force $T_i$ developed at the constrained segments of the rod due to the motion of the sleeves, 
\begin{equation}
    \mathbf{R}_i=-\mathbf{C}_i+T_i \mathbf{b}_i.
\end{equation}

The  strong form of the equations of motion, along with the boundary conditions and the relation between $\bR_i$ and $M_i$, Eq. \eqref{eq:configforce}, are consistent with the derivation reported by Armanini et al. \cite{ARMANINI201982}, as evident from a comparison of Eq. \eqref{panosisstrong}, and \eqref{eq:configforce} with Eq. (15), and Eq. (27) in \cite{ARMANINI201982} respectively.

It is noted that although Eq. \eqref{panosisstrong} is representative of a conservative system,  the expression of the system in a non-material frame entails the contribution of advection terms.
Therefore, since the weak form of the ALE formulation is used to obtain the numerical model in Sect. \ref{sec:discretization}, it is expected that the conservation of energy may be violated when a standard time-stepping algorithm is used. To this purpose, special attention has to be paid to the time integration method for the numerical implementation of the model, as disclosed in Sect. \ref{sec:discretization}.

\subsection{Rod constrained by a sliding sleeve at one end only} \label{sec:one_sleeve}

The case where the rod is constrained by only one sliding sleeve can be obtained by setting $s_2=L$, $\bR_2=0$, and $M_2=0$. It follows also that  the variation $\delta s_2$ vanishes and therefore  Eq. \eqref{eq:gov_eq}$_2$ with $i=2$ becomes trivial, allowing for the reduction of the remaining ones as
\begin{equation} \label{eq:gov_eq2}
     \begin{aligned}
        & \begin{multlined}
            \int_{s_1(t)}^{L} \left\{ \gamma \left[ \frac{\mbox{D}^2 \bx }{\mbox{D} t^2} - \ddot s\ \bx_{,s}  + j^\prime \frac{\mbox{D} \bx }{\mbox{D} t} - \dot s \left( \frac{\mbox D \bx}{\mbox D t} \right)_{\!\!,s} - \bg \right]\cdot \delta\bx + \left[ \gamma\, \dot s\, \left( \frac{\mbox{D} \bx }{\mbox{D} t} - \dot s\ \bx_{,s} \right) + N\,\bx_{,s} \right]\cdot \delta \bx_{,s} \right.
            + B\,\bx_{,ss}\cdot\delta\bx_{,ss} \bigg\} \diff s \\ -\sum\limits_q^Q \bF_q(t)\cdot \bx(t,s_q^*) + \bR_1 \cdot \delta\bx(s_1) + M_1\,\bn_1\cdot\delta\bx_{,s}(s_1) = 0,
        \end{multlined} \\
        & \begin{multlined}
            \left\{ -\left( \frac{\gamma}{2} \dot\bx^2 - \frac{B}{2}\bx_{,ss}^2 + \gamma \bg\cdot \bx \right) + \mathcal T_{C1,s_1} - (\mathcal{T}_{C1,\dot s_{1}})_{,t} - \mathcal V_{C1,s_1} + (\mathcal{V}_{C1,\dot s_{1}})_{,t}\right. 
            - \bR_1 \cdot \bx_{,s}(s_1) - M_1\,\bn_1\cdot\bx_{,ss}(s_1) \bigg\} \delta s_1 = 0.
        \end{multlined}
    \end{aligned}
\end{equation}

\section{Numerical FE scheme for the ALE formulation} \label{sec:discretization}

A FE model is developed for reconstructing the unknown fields of the position $\bx(t,s)$ and of the internal axial force $N(t,s)$ within the time-varying subdomain $\mathcal{N}(t)$ through the use of discrete data.
Similarly to classical FE approaches, the reconstructed fields should fulfill the continuity of the physical fields. However, differently from the  classical Lagrangian formulations, the quantities are defined on the moving mesh in the ALE approach, meaning that the time derivatives obtained from the reconstruction of the time derivatives of the discrete degrees of freedom are in fact \emph{material} time derivatives. 

In order to adopt an adequate reconstruction of the continuous fields $\bx(t,s)$ and $N(t,s)$, the function spaces to which these fields belong must first be defined.
To capture the nonlinear dynamics of an inextensible flexible rod, as in the cases described in Sections \ref{sec:two_sleeve} and \ref{sec:one_sleeve}, $\bx(t,s)$ and $N(t,s)$ should be $\mathsf{C}^1$ and $\mathsf{C}^0$ continuous in the time-varying subdomain $\mathcal N(t)$. Further, since  weak solutions to Eq. \eqref{eq:gov_eq} are sought, the inner products of $\bx(t,s)$, $\bx_{,s}(t,s)$, $\bx_{,ss}(t,s)$, and $N(t,s)$ with their variations (belonging to the same function spaces) need to be defined. Therefore, $\bx$,  $N$ and their variations should belong to the Hilbert spaces $\mathsf{H}^2(\mathcal N)$ and $\mathsf{H}^1(\mathcal N)$ respectively, where the  notation $\mathsf{H}^k(\mathcal N)$ is employed to denote the Sobolev space $\mathsf{W}^{k,2}(\mathcal N)$ over the real numbers, with $\mathsf{H}^0(\mathcal N)\equiv \mathsf{L}^2(\mathcal N)$, and $\mathsf{L}^2(\mathcal N)$ is the space of the square-integrable functions in $\mathcal N$.

To facilitate the reconstruction using finite elements, subdomain $\mathcal N$ is partitioned in $N_{el}$ non-overlapping elements covering the subdomains $\Omega_i=[\widehat s_i, \widehat s_{i+1}]$, with $\widehat s_1 = s_1(t)$ and $\widehat s_{N_{el}+1} = s_2(t)$.
Then, the reconstructed fields (denoted with $\hat{(\,\cdot\,)}$) can be written as
\begin{equation}
    \hat\bx(t,s) = \bH(\hat\sigma(s)) \cdot \hat\bx_i(t), \quad \hat N(t,s) = \bP(\hat\sigma(s)) \cdot \hat \bN_i(t), \quad s\in \Omega_i,
\end{equation}
where the linear map $\hat\sigma(s; i): (\hat s_i, \hat s_{i+1}) \to (0,1)$ is the element-local position parameter, $\bH(\hat\sigma)$, and $\bP(\hat\sigma)$ are the interpolation matrices for the fields $\bx$ and $N$ over $\Omega_i$,
\begin{equation}
    \bH(\hat\sigma) = \left[\ h_1(\hat\sigma)\, \bI\ |\ h_2(\hat\sigma)\, \bI\ |\ h_3(\hat\sigma)\, \bI\ |\ h_4(\hat\sigma)\, \bI\ \right],\quad \bP(\hat\sigma) = \left[\ p_1(\hat\sigma)\ |\ p_2(\hat\sigma)\ \right],
\end{equation}
with $h_k(\hat\sigma), k=1,\dots,4$ being the four two-node Hermite basis functions of $\mathsf{H}^2(\Omega_i)$,  $p_k(\hat\sigma), k=1,2$  the two Lagrange basis functions of $\mathsf{H}^1(\Omega_i)$, and $\bI$ is the identity matrix of appropriate size \cite{hughes2012finite}. 
The vectors $\hat{\bx}_i$, and $\hat{\bN}_i$ collect the degrees of freedom of the discrete reconstruction associated with the $i$-th element. Further, if the entirety of the degrees of freedom is collected in the vectors $\hat\bx(t)$ and $\hat\bc(t)$, where $\hat\bc(t)=[\hat\bN, \bR_1, M_1, \bR_2, M_2]^T$ collects all the degrees of freedom associated with the constraints, the space-discrete governing equations can be written in matrix form as
\begin{equation} \label{eq:discrete}
    \begin{bmatrix}
        \bM_{xx} & 0 \\ 0 & 0
    \end{bmatrix} \cdot 
    \begin{pmatrix}\ddot{\hat{\bx}} \\ \ddot{\hat{\bc}} \end{pmatrix} + 
    \begin{bmatrix}
        \bC_{xx} & 0 \\ 0 & 0
    \end{bmatrix} \cdot 
    \begin{pmatrix}\dot{\hat{\bx}} \\ \dot{\hat{\bc}} \end{pmatrix}
    +
    \begin{bmatrix}
        \bK_{xx} & \bK_{xc}(\hat{\bx}) \\ \bK_{cx}(\hat{\bx}) & 0
    \end{bmatrix} \cdot
    \begin{pmatrix}\hat{\bx} \\ \hat{\bc} \end{pmatrix} = \begin{pmatrix} \bF_x \\ \bF_c \end{pmatrix},
\end{equation}
where
\begin{equation}
\begin{aligned}
    & \bM_{xx} = \mathcal{A}\left(\int_{\hat s_i}^{\hat s_{i+1}} \gamma\, \bH^T \bH \diff s \right), \ 
    \bK_{xx} = \mathcal{A}\left( \int_{\hat s_i}^{\hat s_{i+1}} \left[ B\, \bH_{,ss}^T \bH_{,ss} - \gamma\, \dot s^2\, \bH_{,s}^T \bH_{,s} - \gamma\, \ddot s\,\bH^T \bH_{,s} \right] \diff s \right), \\
    & \bK_{xc} = \mathcal{A}\left( \int_{\hat s_i}^{\hat s_{i+1}} \bH_{,s}^T \bx_{,s} \bP \diff s \right) + \mathcal{A}_{R1}\left( \bH^T(0) \right) + \mathcal{A}_{M1}\left( \bH^T_{,s}(0) \bn_1 \right) + \mathcal{A}_{R2}\left( \bH^T(1) \right) + \mathcal{A}_{M2}\left( \bH^T_{,s}(1) \bn_2 \right),\\
    & \bK_{cx} = \mathcal{A}^T\left( \frac{1}{2} \int_{\hat s_i}^{\hat s_{i+1}} \bP^T \bx_{,s} \bH_{,s} \diff s \right) + \mathcal{A}_{R1}^T\left( \bH(0) \right) + \mathcal{A}_{M1}^T\left( \bn_1^T \bH_{,s}(0) \right) + \mathcal{A}_{R2}^T\left( \bH(1) \right) + \mathcal{A}_{M2}^T\left( \bn_2^T \bH_{,s}(1) \right), \\
    & \bC_{xx} = \mathcal{A}\left(\gamma \int_{\hat s_i}^{\hat s_{i+1}} \left[ j^\prime\, \bH^T \bH + \dot s\ (\bH_{,s}^T \bH - \bH^T \bH_{,s}) \right] \diff s \right), \\
    & \bF_x = \mathcal{A}\left(\gamma \int_{\hat s_i}^{\hat s_{i+1}} \bH^T\, \bg\, \diff s \right) + \sum\limits_i \mathcal{A}_{i}\left( H^T(\hat{\sigma}(s^*_i))\, \bF_{ext} \right) , \\
    & \bF_c = \mathcal{A}\left(\int_{\hat s_i}^{\hat s_{i+1}} \bP^T \diff s \right) + \mathcal{A}_{R1}^T\left( \ba_1 \right) + \mathcal{A}_{R2}^T\left( \ba_2 \right),
\end{aligned}
\end{equation}
and $\mathcal{A}$, $\mathcal{A}_{Ri}$, $\mathcal{A}_{Mi}$ are the appropriate assembly operators correlating local matrices to global degrees of freedom, and the superscript $T$ stands for the transpose  operator.
Finally, this system of equations is augmented by the interface conditions \eqref{eq:gov_eq}$_2$, in order to balance the number of equations and the unknowns $\hat\bx$, $\hat\bc$, and $\bp$.

The Newmark method is used for the time integration of the discretized governing equations \cite{hughes2012finite}, as it provides a stable, simple, and very efficient way to advance from one timestep to the next. Indeed, if the values $\bq^{n}$ and $\bq^{n+1}$ denote the value of the quantity $\bq$ associated with times $t^n$ and $t^{n+1}=t^n+\tau$ respectively (being $\tau>0$ the timestep size), the approximation
\begin{equation} \label{eq:newmark}
    \bq^{n+1} = \bq^n + \tau\, \dot\bq^{n} + \frac{\tau^2}{2} \left[ (1-2\beta_1)\, \ddot\bq^n + 2\beta_1\, \ddot\bq^{n+1} \right], \quad \quad \dot\bq^{n+1} = \dot\bq^n + \tau\,\left[(1-\beta_2)\,  \ddot\bq^n + \beta_2 \, \ddot\bq^{n+1}\right],
\end{equation}
can be used to solve the nonlinear system \eqref{eq:discrete} for accelerations using an iterative process for each timestep; i.e. Newton-Raphson iterations (under a tolerance 10$^{-7}$). The coefficients $\beta_1\in[0,1/2]$ and $\beta_2\in[0,1]$ define the character of the time-integration  \cite{hughes2012finite} and are assumed in the following as $\beta_1=0.255$ and $\beta_2=0.505$. This choice follows from the fact that 
the usually adopted values $\beta_1=0.25$ and $\beta_2=0.5$ (middle point rule), providing in classical problems energy conservation, would instead lead to numerical instabilities in the ALE formulation.

In this case the existence of terms with quadratic dependence on the degrees of freedom introduces terms of $O(\tau^4)$ in the system of equations and thus there exists a lower limit for $\tau$ when the terms $O(\tau^4)$ become insignificant compared to terms of $O(\tau)$ and constants. Then the limitations of the machine precision may lead to inaccuracies in the solution and subsequently the convergence of the time-stepping algorithm suffers. This can be improved significantly by solving for $\left(\hat\bx,\hat\bN\right)^T$ and substituting $\left(\dot{\hat\bx},\dot{\hat\bN}\right)^T$ and $\left(\ddot{\hat\bx},\ddot{\hat\bN}\right)^T$ via Eq. \eqref{eq:newmark}. This approach introduces terms of at most $O(\tau^2)$.

Then, the fully discretized nonlinear equations to be solved at every timestep $n$+1 can be written as 
\begin{equation} \label{eq:full_discrete}
    \begin{aligned}
        & \bbf(\hat{\bx}^{n+1}, \hat{\bc}^{n+1}; \bp^{n+1}) = \begin{bmatrix}
            \frac{1}{\tau^2 \beta_1}\bM_{xx}^{n+1} + \frac{\beta_2}{\tau \beta_1} \bC_{xx}^{n+1} + \bK_{xx}^{n+1} & \bK_{xc}^{n+1} \\
            \bK_{cx}^{n+1} & 0
        \end{bmatrix}
        \begin{pmatrix}
            \hat{\bx}^{n+1} \\ \hat{\bc}^{n+1}
        \end{pmatrix} - 
        \begin{pmatrix}
            \bb_x^{n+1} \\ \bF_c^{n+1}
        \end{pmatrix} = \b0, \\
        & \bg(\hat{\bx}^{n+1}, \hat{\bc}^{n+1}; \bp^{n+1}) = \b0,
    \end{aligned}
\end{equation}
where
\begin{equation}
    \bb_x^{n+1} = \bF_x^{n+1} + \bM_{xx}^{n+1}\cdot\left[ \frac{1}{\tau^2\beta_1} \hat\bx^n + \frac{1}{\tau \beta_1} \dot{\hat\bx}^n + \frac{1-2\beta_1}{2\beta_1} \ddot{\hat\bx}^n \right] + \bC_{xx}^{n+1}\cdot \left[ \frac{\beta_2}{\tau \beta_1} \hat\bx^n - \left( 1 - \frac{\beta_2}{\beta_1}  \right) \dot{\hat\bx}^n - \tau \left(1-\frac{\beta_2}{2\beta_1}\right) \ddot{\hat\bx}^n \right].
\end{equation}

The Jacobian matrix $\bJ$ has to be considered for implementing the Newton-Raphson method. From Eq. \eqref{eq:full_discrete} it follows  that $\bJ$ is a block matrix involving large blocks of null elements. More specifically, for the timestep $n+1$, $\bJ$  has the following form  
\begin{equation}
    \bJ =  
    \begin{bmatrix}
        \bJ_{11} & \bJ_{1p} \\
        \bJ_{p1} & \bJ_{pp}
    \end{bmatrix} 
    = \begin{bmatrix}
        \nabla_{ \hat{\bx}^{n+1}, \hat{\bc}^{n+1}} \bbf & \nabla_{ \hat{\bp}^{n+1} } \bbf \\
        \nabla_{ \hat{\bx}^{n+1}, \hat{\bc}^{n+1}} \bg & \nabla_{ \hat{\bp}^{n+1} } \bg
    \end{bmatrix}
\end{equation}
and the matrix $\bJ_{11}$ can be evaluated easily in the absence of solution dependent forces (i.e. nonlinear viscous damping, friction). Further, being a block matrix,  the form of $\bJ$ can be used to write the inverse $\bJ^{-1}$ as a function of inverses of matrices of smaller dimensions as follows
\begin{equation}
    \bJ^{-1} = 
    \begin{bmatrix}    
        \bJ_{11}^{-1} + \bJ_{11}^{-1}\cdot \bJ_{1p}\cdot \bar\bJ^{-1} \cdot \bJ_{p1}\cdot \bJ_{11}^{-1} & -\bJ_{11}^{-1}\cdot \bJ_{1p}\cdot\bar\bJ^{-1} \\
        -\bar\bJ^{-1}\cdot \bJ_{p1}\cdot \bJ_{11}^{-1} & \bar\bJ^{-1}
    \end{bmatrix},
\end{equation}
where $\bar\bJ$ is the Schur complement of matrix $\bJ$ with respect to block $\bJ_{11}$,
\begin{equation}
    \bar\bJ = \bJ_{pp} - \bJ_{p1}\cdot \bJ_{11}^{-1}\cdot \bJ_{1p}.
\end{equation}

In turn, $\bJ_{11}^{-1}$ can  be calculated using the same technique as
\begin{equation}
    \bJ_{11}^{-1} = 
    \begin{bmatrix}
        \bJ_{xx}^{-1} + \bJ_{xx}^{-1}\cdot \bJ_{xc}\cdot \bar\bJ_{11}^{-1} \cdot \bJ_{cx}\cdot \bJ_{xx}^{-1} & -\bJ_{xx}^{-1}\cdot \bJ_{xn}\cdot\bar\bJ_{11}^{-1} \\
        -\bar\bJ_{11}^{-1}\cdot \bJ_{cx}\cdot \bJ_{xx}^{-1} & \bar\bJ_{11}^{-1}
    \end{bmatrix},
\end{equation}
where $\bar\bJ_{11}$ is the Schur complement of matrix $\bJ_{xx}$
\begin{equation}
    \bar\bJ_{11} = - \bJ_{cx}\cdot \bJ_{xx}^{-1} \cdot \bJ_{xc}.
\end{equation}

The discretized equations are solved numerically using  in-house software developed in Matlab, the source code of which is publicly available \cite{Koutsogiannakis_A_simple_ALE-FEM_2024}.

\section{Validation by case studies and application to instability problems} \label{sec:examples}

Three case studies (\texttt{CS-1}, \texttt{CS-2}, and \texttt{CS-3}) and two instability problems (\texttt{IP-1} and \texttt{IP-2}) are addressed to assess the capability and reliability of the presented ALE formulation. The  time-varying subdomain $\mathcal{N}(t)$ of these five mechanical systems is  discretized through the same number of elements, $N_{el}=32$, while the other numerical data are summarized in Table \ref{Tabledata}. The convergence of the results  with  mesh refinement is also investigated by varying $N_{el}$ for two  case studies (\texttt{CS-2} and \texttt{CS-3}).

\begin{table}[]
    \centering
    \caption{Numerical data and expressions used for the three case studies (\texttt{CS-1}, \texttt{CS-2}, and \texttt{CS-3}) and the two instability problems (\texttt{IP-1} and \texttt{IP-2}).}
    \begin{tabular}{|c|c|c|c|c|c|c|c|c|c|c|}
        \hline
       Data & Units & \texttt{CS-1}  & \texttt{CS-2} & \texttt{CS-3} & \texttt{IP-1} & \multicolumn{4}{c|}{\texttt{IP-2}} \\ \hline
       $L$ & [m] & 1 & 2 & 3 & 2 & \multicolumn{4}{c|}{5} \\
       $B$ & [Nm$^2$] & 2 & 2.8 & 0.15 & 1 & \multicolumn{4}{c|}{1} \\
       $\gamma$ & [kg/m] & 0 & 0.312 & 0.4 & 10$^{-5}$ & \multicolumn{4}{c|}{0.05} \\
       \multirow{2}*{$m$} & \multirow{2}*{[kg]} & \multirow{2}*{1} & \multirow{2}*{-} & \multirow{2}*{-} & 0.999 $m_{tr}$ & \multicolumn{4}{c|}{\multirow{2}*{-}} \\
       & & & & & 1.001 $m_{tr}$ & \multicolumn{4}{c|}{} \\
       $\ell_0$ & [m] & 0.4694 & 1 & 1 & 1 & \multicolumn{4}{c|}{1} \\
       $\theta_1$ & [rad] & $\frac{2\pi}{3}$ & $\frac{\pi}{2}$ & 0 & $\frac{\pi}{4}$ & \multicolumn{4}{c|}{$\omega t$} \\
       $\theta_2$ & [rad] & - & - & 0 & - & \multicolumn{4}{c|}{$\pm\omega t$} \\
       $\omega$ & [rad/s] & - & - & - & - & 0.5 & 0.2 & 0.02 & 10$^{-3}$ \\
      Gravity acc. & - & yes & yes & yes & yes  & \multicolumn{4}{c|}{no}  \\
       Force & [N] & - &  $\sin(4\pi\, t)$ & - & -  & \multicolumn{4}{c|}{-}  \\
        Dissipation & - & - & - &  c=\{0,1\}Ns/m & $\zeta$=0.025, $\mu=0.15$ & \multicolumn{4}{c|}{-}  \\
       $\tau$ & [s] & 10$^{-4}$ & 10$^{-4}$ & 10$^{-3}$ & 10$^{-5}$ &  \multicolumn{3}{c|}{$2\cdot 10^{-3}$} &  0.1 \\ \hline
    \end{tabular}
    \label{Tabledata}
\end{table}

\subsection{Case studies}

\begin{figure}[!h]
    \centering
    \includegraphics[width=160mm]{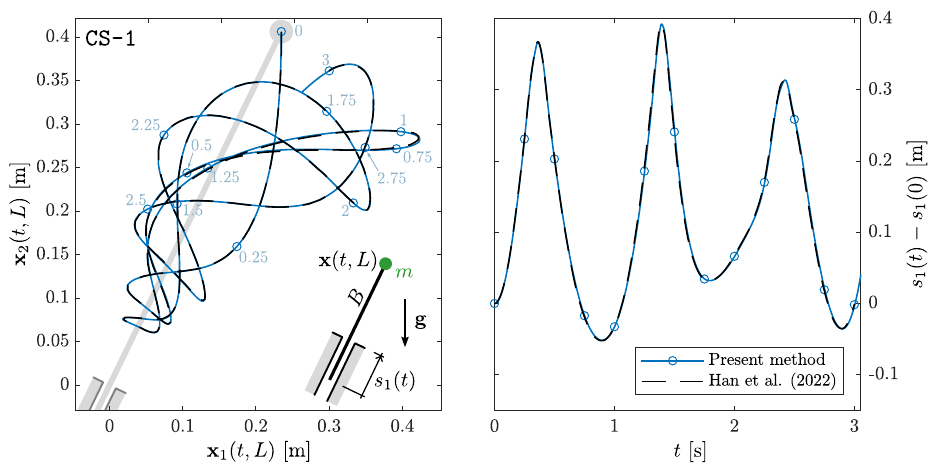}
    \caption{ (Left) Trajectory of the lumped mass $m$ attached at the curvilinear coordinate $s=L$ and (Right) evolution in time of  the insertion length difference $s_1(t)-s_1(0)$ for  \texttt{CS-1}. Results obtained with the present method (solid blue line) are superimposed to those by Han \cite{han2022configurational} (dashed black line). Specific position at  time $t=(0.25j)$s ($j=1,...,12$) is highlighted  along the trajectory.}
    \label{fig:shilei_compare_massless}
\end{figure}

\paragraph{\texttt{CS-1}:} A massless elastic rod has a lumped mass $m$ attached at one end ($s=L$) and is constrained at the other end by a sliding sleeve with constant inclination $\theta_1=2\pi/3$. A  gravity acceleration field is present, opposite to the $x_2$ axis and of magnitude $g=9.81$m/s$^2$. The system evolution is analyzed for the 3s just after the release time $t=0$ from initial rest conditions, where the rod has an initial external length  $\ell_0 = L-s_1(0)$. 
The results from the numerical integration are reported as the  trajectory of the lumped mass in Fig. \ref{fig:shilei_compare_massless}(left) with a blue line marked with circles and as the evolution of the configuration parameter $s_1(t)$ in Fig. \ref{fig:shilei_compare_massless}(right). Results from the same system obtained from a different ALE formulation proposed by Han \cite{han2022configurational} are also reported in  Fig. \ref{fig:shilei_compare_massless} with a dashed black line, showing  excellent accordance between the two ALE formulations.

\begin{figure}[!h]
    \centering
    \includegraphics[width=90mm]{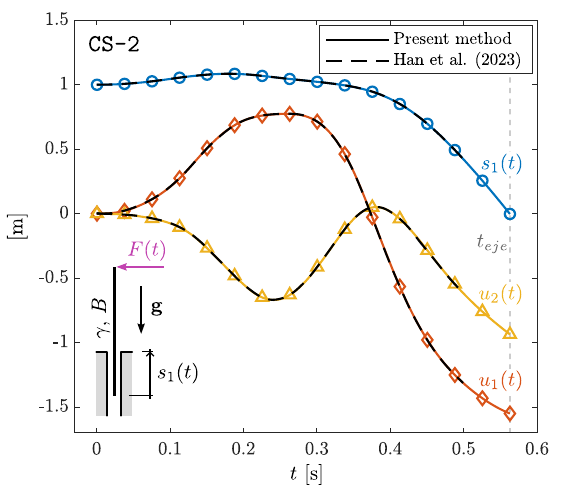}
    \caption{Evolution in time $t$ of the displacement components $u_1(t)$ (red) and $u_2(t)$ (yellow)  of the unconstrained rod's end ($s=L$) and  of the insertion length $s_1(t)$ (blue) for $\texttt{CS-2}$.  The system evolution ends with a  final ejection of the rod at $t=t_{eje}\approx 0.563$s.  Results obtained from the present method (continuous lines) are perfectly superimposed to those by Han and Bachau \cite{han2023configurational} (dashed lines). }
    \label{fig:shilei_compare_mass}
\end{figure}

\paragraph{\texttt{CS-2}:} Mass convection effects are activated by considering a rod with a non-null linear mass density $\gamma$. 
 The rod is constrained by one sliding sleeve with constant inclination $\theta_1=\pi/2$ and loaded, in addition to a gravity field as for \texttt{CS-1},   through a time-harmonic force $F(t)=\sin(4\pi\, t)$ acting at the curvilinear coordinate $s=L$ with constant direction, parallel to the $x_1$ axis. Assuming  initial rest conditions and an initial external length $\ell_0=1$m, the evolution until final ejection of the rod (occurring at $t=t_{eje}\approx 0.563$s) obtained from the present method is reported in Fig. \ref{fig:shilei_compare_mass} for the displacement components  $u_1(t) = x_1(t,L)-x_1(0,L)$ (solid red) and  $u_2(t) = x_2(t,L)-x_2(0,L)$ (solid yellow) at the edge  $s=L$ and for the configuration parameter  $s_1(t)$ (solid blue). These results are validated through comparison with the numerical results available  in   \cite{han2023configurational}, also included in Fig. \ref{fig:shilei_compare_mass}  with dashed black lines. 
It is noted that the system becomes unconstrained for $t> t_{eje}$ and therefore its dynamics  can be solved by standard methods afterwards.

\begin{figure}[!h]
    \centering
    \includegraphics[width=160mm]{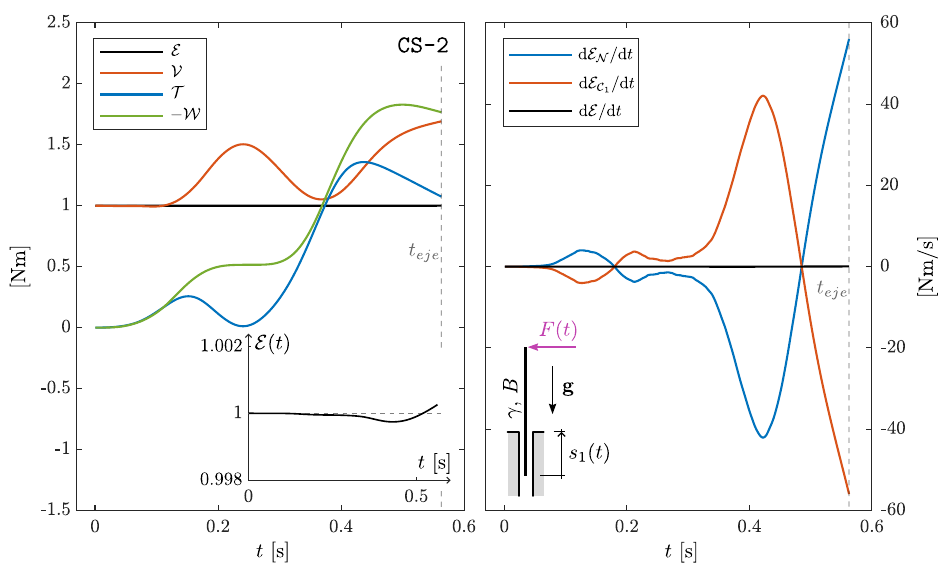}
    \caption{  (Left) Total $\mathcal{E}$ (black), potential $\mathcal{V}$ (red), and kinetic  $\mathcal{T}$ (blue) energies, and the negative of the work of the external force $\mathcal{W}$ (green) as a function of time $t$ for \texttt{CS-2}, until $t=t_{eje}\approx 0.563$s when the final ejection of the rod occurs. 
    A magnification of   the total energy $\mathcal{E}$ curve is also included in the inset to  highlight the violation of the conservation of energy principle by the numerical integration scheme. (Right) Time derivative of the total energies $\mathcal{E}_\mathcal{N}$ (blue) and $\mathcal{E}_{\mathcal{C}_1}$ (red) respectively associated with  the varying subdomains $\mathcal N(t)$ and $\mathcal C_1(t)$. The time derivative of these two quantities are of equal measure and opposite as shown by the black line representing the derivative in time of the total energy $\mathcal{E}$ of the whole system.}
    \label{fig:energy_derivative}
\end{figure}

The kinetic $\mathcal{T}$ and potential $\mathcal{V}$ energies,  the negative of the external work $\mathcal{W}$ of  the time-harmonic force, and the total energy $\mathcal E=\mathcal T + \mathcal V-\mathcal W$ of the whole system is 
reported in Fig. \ref{fig:energy_derivative}(left) 
as a function of time. It is observed that, although the energy appears  conserved,  advection effects inherent to the ALE formulation introduce a deviation in time of the total energy $\mathcal E(t)$ from its initial value $\mathcal E(0)$ during fast dynamics transients, as shown in the inset (drawn by magnifying the vertical axis). To overcome such a small violation of energy conservation, an upwind biased numerical scheme can potentially be used.
The results for the energies (Fig. \ref{fig:energy_derivative}, left) are complemented by the time derivative of the total energies $\mathcal E_{\mathcal{N}}=\mathcal T_{\mathcal{N}} + \mathcal V_{\mathcal{N}}-\mathcal W$ (blue) and  $\mathcal E_{\mathcal{C}_1}=\mathcal T_{\mathcal{C}_1} + \mathcal V_{\mathcal{C}_1}$ (red), respectively associated to the two subdomains $\mathcal{N}(t)$ and $\mathcal{C}_1(t)$ in Fig. \ref{fig:energy_derivative}(right). In addition to these two curves, a third curve in black representing $\diff \mathcal{E}/\diff t = \diff(\mathcal{E}_{\mathcal{N}}+\mathcal{E}_{\mathcal{C}_1})/\diff t$ is reported,  confirming that, no significant energy change occurs for the whole system, although each rod's subportion displays a varying energy in time.

Finally, it is noted that the reported curves are obtained by assuming a unit value for the initial total energy, $\mathcal E(0)=1$Nm.

\begin{figure}[!h]
    \centering
    \includegraphics[width=160mm]{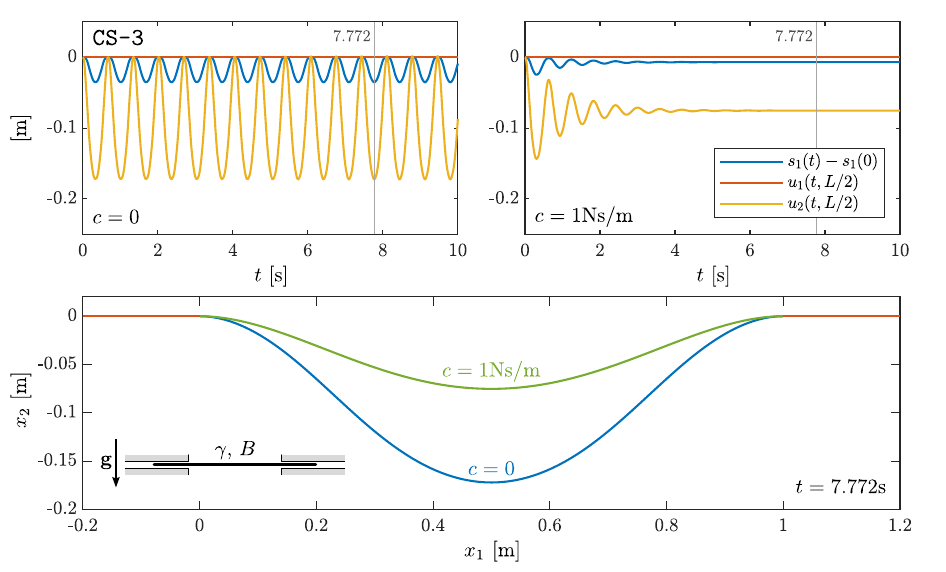}
    \caption{(Top) Evolution in time $t$ of $s_1(t)-s_1(0)$ (blue) and of the displacement components $u_1(t)=x_1(t,L/2)-x_1(0,L/2)$ (red), $u_2(t)=x_2(t,L/2)-x_2(0,L/2)$ (yellow) associated to the rod's mid point ($s=L/2$) for \texttt{CS-3}. Two cases are considered, one with null dissipation ($c=0$, top left) and the other with a non-null dissipation ($c=1$Ns/m, top right). (Bottom) Deformed shape of the rod at time $t=7.772$s for the undissipated  ($c=0$, blue curve) and dissipated ($c=1$Ns/m, green curve) systems.}
    \label{fig:two_sleeve_timeseries}
\end{figure}

\paragraph{\texttt{CS-3}:} A rod with non-null mass density $\gamma$ within a gravity field, defined as in \texttt{CS-1}, is constrained by two  sliding sleeves aligned along the $x_1$ axis and at a distance  $(\ba_2-\ba_1)\cdot\be_1=1$m from each other, where $\be_1$ is the unit vector parallel to the $x_1$ axis.
Dissipation is introduced as a distributed transverse loading along the subdomain $\mathcal{N}(t)$ as
\begin{equation}
    \bF_{dissip} = - c\, \bv_{\perp}(t)
\end{equation}
where $ \bv_\perp = \dot\bx - \bx_{,s}\, (\bx_{,s}\cdot \dot\bx )$ is the component of the velocity $\dot\bx$ perpendicular to the rod and $c\geq0$ is the dissipation coefficient.
From an initial rest condition, the system evolves with oscillations of decaying amplitude whenever $c>0$. The evolution is displayed through the values of $s_1(t)-s_1(0)$ (blue),  $u_1(t)=x_1(t,L/2)-x_1(0,L/2)$ (red), $u_2(t)=x_2(t,L/2)-x_2(0,L/2)$ (yellow) in Fig. \ref{fig:two_sleeve_timeseries} for the first 10 s just after the release time for the conservative system ($c=0$, top left) and the non-conservative one ($c=1$Ns/m, top right).
The deformed configuration at $t=7.772$s  is shown in Fig. \ref{fig:two_sleeve_timeseries}(bottom) for the subdomain $\mathcal{N}$ of these two systems ($c=0$ blue, $c=1$Ns/m green)
 along with part of the fully constrained portions $\mathcal{C}_1$ and $\mathcal{C}_2$. The  time $t=7.772$s is selected as providing the (periodic) maximum deflection of the rod for the  conservative system ($c=0$).

\subsection{Convergence of mesh, timestep size and inextensibility constraint}

Case studies \texttt{CS-2} and the undissipated version of \texttt{CS-3} are further investigated to analyze the solution  convergence  by varying the number of elements through the sequence $N_{el}=2^n, n=1,..,6$, under a constant  timestep $\tau=5\cdot10^{-4}$s. The $\mathsf{L}^2$ error estimates for the trajectory of the deformed shape and the configuration parameter vector $\bp$, with respect to corresponding values for the case with $N_{el}=128$ (considered here as the target  solution) are evaluated as
\begin{equation}\label{err_est1}
    \epsilon_{\bx} = \left[ \int_{0}^{1} \left[\bx_{N_{el}}(t_c,s(t_c,\sigma)) - \bx_{128}(t_c,s(t_c,\sigma))\right]^2 \diff \sigma  \right]^{\frac{1}{2}}, \qquad
    \epsilon_{\bp} = ||\bp_{N_{el}}(t_c) - \bp_{128}(t_c)||,
\end{equation}
and, similarly, for the time derivatives the error estimates  as
\begin{equation}\label{err_est2}
    \epsilon_{\dot\bx} = \left[  \int_{0}^{1} \left[\dot\bx_{N_{el}}(t_c,s(t_c,\sigma)) - \dot\bx_{128}(t_c,s(t_c,\sigma))\right]^2 \diff \sigma  \right]^{\frac{1}{2}}, \qquad
    \epsilon_{\dot\bp} = ||\dot\bp_{N_{el}}(t_c) - \dot\bp_{128}(t_c)||,
\end{equation}
where $t_c$ is the reference time at which the convergence is assessed.

\begin{figure}[!h]
    \centering
    \includegraphics[width=160mm]{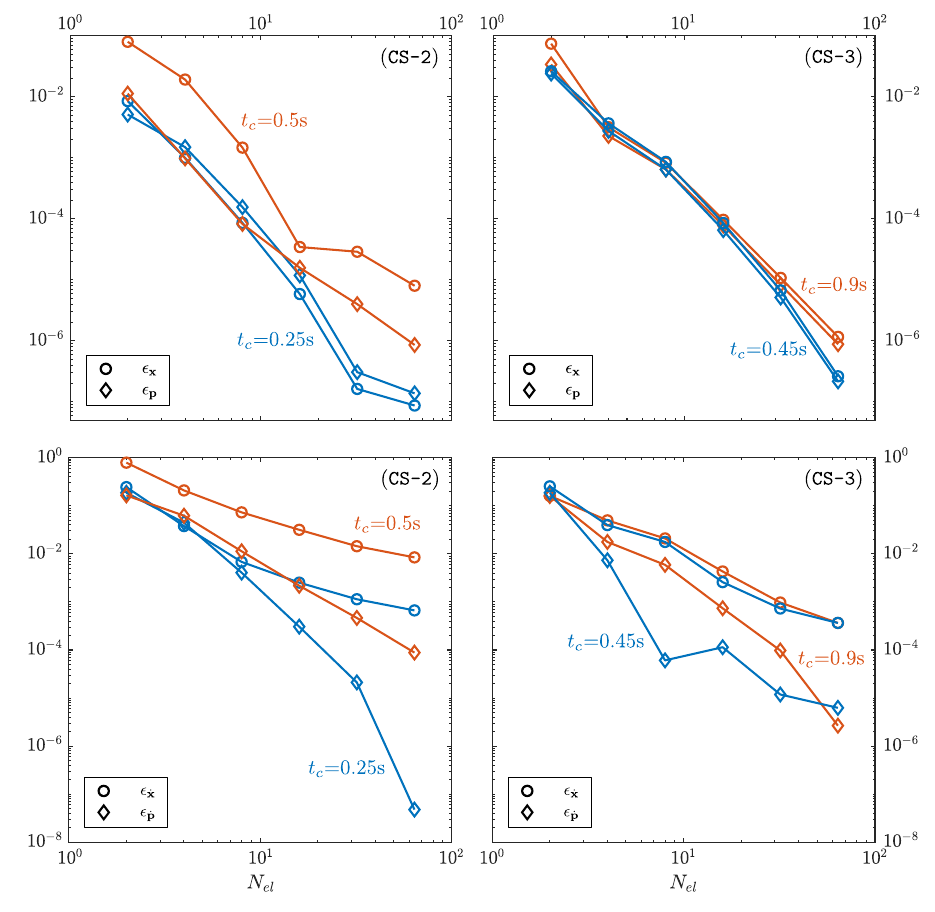}
    \caption{(Left) Convergence analysis for \texttt{CS-2} at the two reference times $t_c=0.25$s (blue lines) and $t_c=0.5$s (red lines). The top part shows the error estimates for the trajectory $\bx$ (circles) and  the configuration parameter vector $\bp$ (diamonds) at increasing   number of elements $N_{el}$ under a timestep  $\tau=5\cdot10^{-4}$s, while the lower part shows the error etsimates for their derivatives. (Right) As for left, but for the undissipated version of \texttt{CS-3} at  $t_c=0.45$s (blue lines) and $t_c=0.9$s (red lines). }
    \label{fig:converge}
\end{figure}

The convergence of the error estimates 
is shown in 
Fig. \ref{fig:converge}
 for \texttt{CS-2} (left) and for \texttt{CS-3} (right).
Two  reference times are considered for each case study, $t_c=\{0.25,0.5\}$s for  \texttt{CS-2} and $t_c=\{0.45,0.9\}$s
for \texttt{CS-3}, distinguished by using two different colors.
Error estimates are reported for $\epsilon_{\bx}$ and $\epsilon_{\bp}$ (top) and  for $\epsilon_{\dot\bx}$ and $\epsilon_{\dot\bp}$ (bottom), evaluated through
Eqns \eqref{err_est1} and \eqref{err_est2}. The error estimates related to the trajectory $\mathbf{x}$ (configuration parameter vector $\mathbf{p}$) and to its time derivative are denoted with circle (diamond) markers. The Figure shows that the errors of the solution obtained by the present method decrease exponentially with the number of elements, thus achieving convergence.

\subsection{Application to instability problems and prediction of critical conditions}

The present method is finally tested through its application to two different instability problems \texttt{IP-1} and \texttt{IP-2} for assessing if the evaluated critical values of the input parameters match the respective predictions provided in \cite{ARMANINI201982} and in 
\cite{cazzolli2024elasticasling}. These examples are important, as they are representative of cases where the loss of stability of an equilibrium configuration leads to fast dynamics. Numerical data considered for  \texttt{IP-1} and \texttt{IP-2} are summarized in Table \ref{Tabledata}.

\paragraph{\texttt{IP-1}:} This problem is very similar to \texttt{CS-1} with  exception for the values of the input data $L$, $B$, $\theta_1$, and the presence of a small self-weight $\gamma$ in order to include dissipation in the system at two end points of the domain $\mathcal{N}(t)$ within the present formulation. More specifically, dissipation is included through:
\begin{itemize}
    \item  a concentrated viscous force $\bF_{visc}(t)$ applied at the lumped mass curvilinear coordinate $s=L$
\begin{equation}
    \bF_{visc}(t) = - 2\,\zeta \sqrt{ \frac{3 m B}{(L-s_1(t))^3} } \,\dot\bx(t,L),
\end{equation}
where $\zeta\geq 0$ is the dimensionless viscous dissipation parameter;
\item a concentrated Coulomb friction force $\bF_{fr}(t)$ at the sliding sleeve exit  ($s=s_1(t)$),  acting parallel to the sliding direction $\bb_1$ and oppositely to the sliding velocity $\dot{s}_1(t)$, defined as
\begin{equation}
    \bF_{fr}(t) = - \mu\ |\bR(t)\cdot \bn|\ \text{sgn}(\dot s_1(t)) \, \bb_1,
\end{equation}
where $\mu\geq 0$ is the friction coefficient, $\bR(t)\cdot\bn$  is the reaction force component perpendicular to the sliding sleeve at its exit, Eq.(\ref{eq:configforce}), and $\text{sgn}(\cdot)$ is the sign function. For the purposes of the simulation, the friction force $\bF_{fr}(t)$ is approximated through the smooth function
\begin{equation}
    \bF_{fr}(t) = - \mu\ \sqrt{ (\bR(t)\cdot \bn)^2 }\ \frac{\dot s_1(t)}{\sqrt{\dot s_1^2(t) + \varepsilon}}  \, \bb_1,
\end{equation}
where $\varepsilon$ is a small quantity,  selected as $\varepsilon=2\cdot 10^{-6}$m$^2$/s$^2$.
\end{itemize}

According to \cite{ARMANINI201982}, as the result of the dynamic evolution after the release of the undeformed rod from a state of rest,  two different final states are expected: either the rod is completely injected inside the sleeve or it is completely ejected from the sleeve. 
For given values of the bending stiffness $B$, dissipation parameters ($\zeta$, $\mu$), and of the initial external length $\ell_0$, each one of the two different final states is associated to a connected  region within the $m$--$\theta_1$ plane, and the transition between these two behaviours is provided by the monotonic curve $m_{tr}(\theta_1)$.
For the  angle $\theta_1$ under consideration, the transition value for the lumped mass is found in \cite{ARMANINI201982} under the assumption of null linear mass density ($\gamma=0$) to be given by $m_{tr}(\theta_1=\pi/4)\approx 0.184098$kg. 

\begin{figure}[!h]
    \centering
    \includegraphics[width=\textwidth]{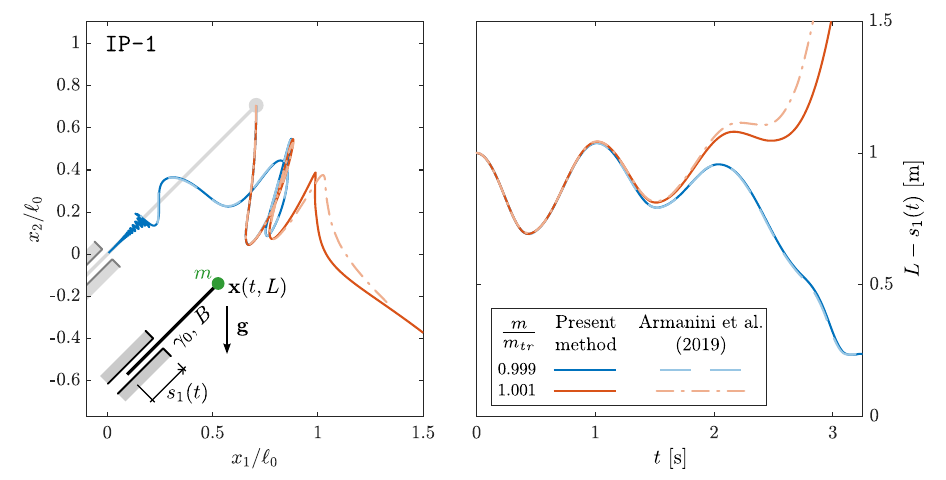}
    \caption{(Left) Trajectory of the lumped mass $m$ attached at the rod's end $s=L$. (Right) Evolution in time of the external rod's length, $L-s_1(t)$. The curves are reported for $m=0.999m_{tr}$ (blue) and for $m=1.001m_{tr}$ (red), where continuous lines represent the result from the present ALE method while the dashed lines from \cite{ARMANINI201982}. The  difference in the response between the two models for the value $m=1.001 m_{tr}$ is due to the high sensitivity of the system near-instability conditions and arises because of the presence of a small distributed inertia $\gamma$ in the present method, which was considered null in \cite{ARMANINI201982}.}
    \label{fig:armanini_transition}
\end{figure}

The results of the present ALE method are shown in Fig. \ref{fig:armanini_transition} for two different lumped mass values $m=\{0.999,\, 1.001\}\, m_{tr}$ attached to a rod with a linear mass density  $\gamma=1.0864 \cdot 10^{-4}$kg/m and are compared with the semi-analytical solution based on the Euler's elastica \cite{ARMANINI201982} (obtained by assuming $\gamma=0$). More specifically, the trajectories of the lumped mass $m$ and the time evolution of the length of the rod outside the sleeve $L-s_1(t)$ are reported on the left and on the right of Fig. \ref{fig:armanini_transition} respectively. The curves corresponding to $m=0.999\, m_{tr}$ are reported with solid blue lines and to $m=1.001\, m_{tr}$ with solid red lines, and the results from the present ALE method are reported as continuous while those from \cite{ARMANINI201982} as dashed lines. 

The discrepancy in the lumped mass trajectory $\mathbf{x}(L,t)$ and in the external rod's length, $L-s_1(t)$, which can be observed for the case $m=1.001\, m_{tr}$, is inherent to the  effect of the  distributed mass $\gamma$ in the vicinity from above of the  mass value $m$ to $m_{tr}$, which is very small ($\gamma=2.17\cdot 10^{-4}\, m_{tr}/L$) in the present ALE-FE  model, but null  in the elastica-based model by Armanini et al. \cite{ARMANINI201982}. 
Indeed,  by definition of critical condition, the response of the rod becomes very sensitive near the transition value $m_{tr}$ and as a consequence even a small difference in the model (such as considering a small, but non-null, distributed inertia) realizes noticeable differences in the results. However, despite this  difference in the trajectory, the final state of the system is correctly predicted  in both cases.

\begin{figure}[!h]
    \centering
    \includegraphics[width=\textwidth]{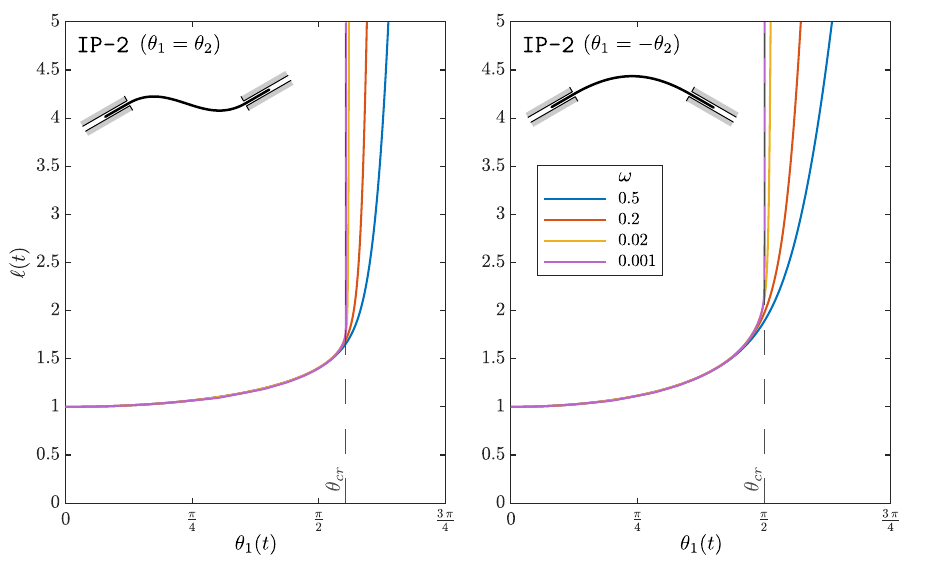}
    \caption{External length $\ell(t)=s_2(t)-s_1(t)$ during the monothonic increase of $\theta_1(t)$ at a constant rate $\omega=\{0.001, 0.02, 0.2, 0.5\}$rad/s,  shown with purple, yellow, red,  and blue curves respectively. 
    Two loading conditions are reported, skewsymmetric ($\theta_1=\theta_2$, left) and symmetric ($\theta_1=-\theta_2$, right).
     The critical angle $\theta_{cr}$ for the uncontrolled ejection evaluated semi-analytically under  quasi-static conditions in \cite{cazzolli2024elasticasling} is shown with a vertical black line, corresponding to  $\theta_{cr} = 1.7378$ (left) and to  $\theta_{cr} = \pi/2$ (right). }
    \label{fig:rot_snap}
\end{figure}

\paragraph{\texttt{IP-2}:} This problem is inspired by the elastica sling system recently presented in \cite{cazzolli2024elasticasling}, where an elastic rod is constrained by two sliding sleeves whose position and inclination can be independently controlled in time. Within a quasi-static setting, critical conditions for the inclination of the sliding sleeve have been  found, for which an indefinite ejection of the rod from the constraint is realized, due to the loss of equilibrium stability.
This mechanical system is treated here within a dynamic framework by taking into account a non-null linear mass density $\gamma$.
Without loss of generality, the sliding-sleeves are rotated around their exit points at a constant rate  $\theta_1=\omega t$ and $\theta_2=\pm\omega t$, where the $+$ ($-$) sign realizes a skewsymmetric (symmetric) loading condition. The critical condition within a quasi-static setting is found in \cite{cazzolli2024elasticasling} to be associated to $\theta_1=\theta_{cr}$ with $\theta_{cr}\approx1.7378$rad when $\theta_1=\theta_2$ and $\theta_{cr}=\pi/2$ when $\theta_1=\theta_2$. 
Results from the analysis performed with the present method at different values of the angular velocity $\omega = \{ 10^{-3}, 0.02, 0.2, 0.5 \}$rad/s are reported through the external rod's length $\ell(t) = s_2(t)-s_1(t)$ versus the input rotation $\theta_1(t)$ in  Fig. \ref{fig:rot_snap}  with purple, yellow, red, and blue lines, respectively. 

Results for the skewsymmetric  loading condition are shown in  Fig. \ref{fig:rot_snap}(left) and for the symmetric one in Fig. \ref{fig:rot_snap}(right). The critical angle $\theta_{cr}$ obtained semi-analytically within a quasi-static setting in \cite{cazzolli2024elasticasling} is also included as a reference as a vertical dashed black line.
In line with previous observations on snapping mechanisms \cite{cazzolli2019snapping,liu2021mechanics,huang2024dominating}, according to which  \emph{the higher the angular velocity, the greater the delay}, the present results confirm  the  delay of the instability due to inertial effect for  the elastica sling problem under both loading conditions. On the other hand, the critical condition evaluated within the quasi-static setting in \cite{cazzolli2024elasticasling} is recovered for very small angular velocity, $\omega=10^{-3}$rad/s.

\section{Conclusion}

An ALE-FE model has been  developed for solving the planar nonlinear dynamics of an elastic rod constrained at both ends by sliding sleeves. For this purpose, the mathematical formulation of the mechanics of time variable domains is presented and the first variation of functionals on time-variable domains is derived. Using a position-based description of the rod, the governing equations of the sliding sleeve-rod system are obtained in both the weak and the strong form. The model is based on the idea of a moving mesh  defined by an underlying time-variable map, from material coordinates to non-material coordinates. The ALE-FE  model is derived by means of a spatial discretization of the relevant fields and the use of the Newmark method for the integration in time. Finally, the numerical model is used to solve some reference case studies and to   accurately predict the critical conditions in instability problems, where critical conditions and fast dynamics govern the performance of proposed solutions.

Although a quantitative comparison would require a fair implementation and execution of the methods to be compared, from a qualitative perspective the developed model appears to have the advantage of being faster and easier to implement than the methods currently available. A minor weakness is that, by using the proposed Newmark integration in time, the system energy  is only approximately conserved during fast dynamic transients, due to the advection effects. However, it is expected that this problem could be overcome by the implementation of an upwind biased integration scheme, which is left for future research.

The proposed ALE-FE model, whose source code is made available, provides a powerful yet simple tool to analyze variable-domain mechanical systems and, in turn, to design novel mechanisms for soft actuation, energy harvesting and wave mitigation. It also represents a first step towards an extension for easily developing novel solvers for the three-dimensional dynamics of flexible one- and two-dimensional systems with varying domains.

\paragraph{CRediT authorship contribution statement.}  
P. Koutsogiannakis: Conceptualization; Methodology; Software; Visualization; Validation; Formal analysis; Investigation; Data Curation; Writing - Original Draft; Writing - Review and Editing.
T. Papathanasiou: Methodology; Formal analysis; Supervision; Writing - Original Draft; Writing - Review and Editing.
F. Dal Corso: Conceptualization; Methodology; Formal analysis; Resources; Supervision; Project administration; Writing - Original Draft; Writing - Review and Editing;  Funding acquisition. 

\paragraph{Acknowledgements.} The authors gratefully acknowledge financial support from the European Research Council (ERC)
under the European Union’s Horizon Europe research and innovation programme, Grant agreement No. 
ERC-ADG-2021-101052956-BEYOND. 
The authors also acknowledge the Italian Ministry of Education, Universities and Research (MUR) in the framework of the project DICAM-EXC (Departments of Excellence 2023-2027, grant L232/2016).
 The methodologies developed in the present work fall within the aims of the GNFM (Gruppo Nazionale per la Fisica Matematica) of the INDAM (Istituto Nazionale di Alta Matematica).

\section*{References}
\printbibliography[heading=none]

\end{document}

%% file: preamble.tex
\usepackage[top=1.5in,right=0.8in,left=0.8in,bottom=1.0in]{geometry}
 
\usepackage{graphics,graphicx}
\usepackage{subfig}
\usepackage{float}
\usepackage{booktabs}
\usepackage{longtable}
\usepackage{supertabular}
\usepackage{tabularx}
\usepackage{multirow}
\usepackage{array}
\usepackage{tabu}
\usepackage{mathtools}
\usepackage{hyperref}
\usepackage{cancel}

\usepackage[ruled,vlined]{algorithm2e}

\usepackage{amsmath}
\usepackage{amsfonts}
\usepackage{amssymb}
\usepackage{siunitx} 

\usepackage{enumerate}
\usepackage{sectsty}
\usepackage[affil-it,auth-sc]{authblk}
\usepackage{color}
\usepackage[dvipsnames]{xcolor}

\sectionfont{\normalsize}
\subsectionfont{\normalsize}

\usepackage{tikz}
\usetikzlibrary{arrows,decorations,backgrounds,shapes, snakes}
\usetikzlibrary{positioning}
\usetikzlibrary{matrix} 
\usepackage{varwidth}
\usepackage[most]{tcolorbox}
\usetikzlibrary{shapes.geometric,arrows.meta,decorations.markings}

%% file: definitions.tex
\newcommand{\singlespace}{\baselineskip=12pt\lineskiplimit=0pt\lineskip=0pt}
\def\ds{\displaystyle}

\tikzstyle{every picture}+=[remember picture]

\newcommand{\beq}{\begin{equation}}
\newcommand{\eeq}{\end{equation}}
\newcommand{\lb}{\label}
\newcommand{\ph}{\phantom}
\newcommand{\beqar}{\begin{eqnarray}}
\newcommand{\eeqar}{\end{eqnarray}}
\newcommand{\barr}{\begin{array}}
\newcommand{\earr}{\end{array}}
\newcommand{\jump}{\parallel}
\newcommand{\Ehat}{\hat{E}}
\newcommand{\That}{\hat{\bf T}}
\newcommand{\Ahat}{\hat{A}}
\newcommand{\chat}{\hat{c}}
\newcommand{\shat}{\hat{s}}
\newcommand{\khat}{\hat{k}}
\newcommand{\muhat}{\hat{\mu}}
\newcommand{\mc}{M^{\scriptscriptstyle C}}
\newcommand{\mei}{M^{\scriptscriptstyle M,EI}}
\newcommand{\mec}{M^{\scriptscriptstyle M,EC}}
\newcommand{\hbeta}{{\hat{\beta}}}
\newcommand{\rec}[2]{\left( #1 #2 \ds{\frac{1}{#1}}\right)}
\newcommand{\rep}[2]{\left( {#1}^2 #2 \ds{\frac{1}{{#1}^2}}\right)}
\newcommand{\derp}[2]{\ds{\frac {\partial #1}{\partial #2}}}
\newcommand{\derpn}[3]{\ds{\frac {\partial^{#3}#1}{\partial #2^{#3}}}}
\newcommand{\dert}[2]{\ds{\frac {d #1}{d #2}}}
\newcommand{\dertn}[3]{\ds{\frac {d^{#3} #1}{d #2^{#3}}}}
\newcommand{\ct}{\captionof{table}}
\newcommand{\cf}{\captionof{figure}}

\def\c{{\circ}}
\def\bob{{\, \underline{\overline{\otimes}} \,}}
\def\ob{{\, \underline{\otimes} \,}}
\def\scalp{\mbox{\boldmath$\, \cdot \, $}}
\def\gdp{\makebox{\raisebox{-.215ex}{$\Box$}\hspace{-.778em}$\times$}}
\def\daa{\makebox{\raisebox{-.050ex}{$-$}\hspace{-.550em}$: ~$}}
\def\mK{\mbox{${\mathcal{K}}$}}
\def\cK{\mbox{${\mathbb {K}}$}}

\def\Xint#1{\mathchoice
   {\XXint\displaystyle\textstyle{#1}}%
   {\XXint\textstyle\scriptstyle{#1}}%
   {\XXint\scriptstyle\scriptscriptstyle{#1}}%
   {\XXint\scriptscriptstyle\scriptscriptstyle{#1}}%
   \!\int}
\def\XXint#1#2#3{{\setbox0=\hbox{$#1{#2#3}{\int}$}
     \vcenter{\hbox{$#2#3$}}\kern-.5\wd0}}
\def\ddashint{\Xint=}
\def\fpint{\Xint=}
\def\dashint{\Xint-}
\def\cpvint{\Xint-}
\def\intl{\int\limits}
\def\cpvintl{\cpvint\limits}
\def\fpintl{\fpint\limits}
\def\ointl{\oint\limits}
\def\bA{{\bf A}}
\def\ba{{\bf a}}
\def\bB{{\bf B}}
\def\bb{{\bf b}}
\def\bc{{\bf c}}
\def\bC{{\bf C}}
\def\bD{{\bf D}}
\def\bE{{\bf E}}
\def\be{{\bf e}}
\def\bbf{{\bf f}}
\def\bF{{\bf F}}
\def\bG{{\bf G}}
\def\bg{{\bf g}}
\def\bi{{\bf i}}
\def\bI{{\bf I}}
\def\bH{{\bf H}}
\def\bJ{{\bf J}}
\def\bK{{\bf K}}
\def\bL{{\bf L}}
\def\bM{{\bf M}}
\def\bN{{\bf N}}
\def\bn{{\bf n}}
\def\bm{{\bf m}}
\def\b0{{\bf 0}}
\def\bO{{\bf O}}
\def\bo{{\bf o}}
\def\bX{{\bf X}}
\def\bx{{\bf x}}
\def\bP{{\bf P}}
\def\bp{{\bf p}}
\def\bQ{{\bf Q}}
\def\bq{{\bf q}}
\def\bR{{\bf R}}
\def\bS{{\bf S}}
\def\bs{{\bf s}}
\def\bT{{\bf T}}
\def\bt{{\bf t}}
\def\bU{{\bf U}}
\def\bu{{\bf u}}
\def\bv{{\bf v}}
\def\bV{{\bf V}}
\def\bw{{\bf w}}
\def\bW{{\bf W}}
\def\by{{\bf y}}
\def\bz{{\bf z}}
\def\T{{\bf T}}
\def\Te{\textrm{T}}
\def\Id{{\bf I}}
\def\bxi{\mbox{\boldmath${\xi}$}}
\def\balpha{\mbox{\boldmath${\alpha}$}}
\def\bbeta{\mbox{\boldmath${\beta}$}}
\def\bepsilon{\mbox{\boldmath${\epsilon}$}}
\def\bvarepsilon{\mbox{\boldmath${\varepsilon}$}}
\def\bomega{\mbox{\boldmath${\omega}$}}
\def\bphi{\mbox{\boldmath${\phi}$}}
\def\bsigma{\mbox{\boldmath${\sigma}$}}
\def\bfeta{\mbox{\boldmath${\eta}$}}
\def\bDelta{\mbox{\boldmath${\Delta}$}}
\def\btau{\mbox{\boldmath $\tau$}}
\def\tr{{\rm tr}}
\def\dev{{\rm dev}}
\def\div{{\rm div}}
\def\Div{{\rm Div}}
\def\Grad{{\rm Grad}}
\def\grad{{\rm grad}}
\def\Lin{{\rm Lin}}
\def\Sym{{\rm Sym}}
\def\Skw{{\rm Skew}}
\def\abs{{\rm abs}}
\def\Re{{\rm Re}}
\def\Im{{\rm Im}}
\def\capB{\mbox{\boldmath${\mathsf B}$}}
\def\capC{\mbox{\boldmath${\mathsf C}$}}
\def\capD{\mbox{\boldmath${\mathsf D}$}}
\def\capE{\mbox{\boldmath${\mathsf E}$}}
\def\capG{\mbox{\boldmath${\mathsf G}$}}
\def\tcapG{\tilde{\capG}}
\def\capH{\mbox{\boldmath${\mathsf H}$}}
\def\capK{\mbox{\boldmath${\mathsf K}$}}
\def\capL{\mbox{\boldmath${\mathsf L}$}}
\def\capM{\mbox{\boldmath${\mathsf M}$}}
\def\capR{\mbox{\boldmath${\mathsf R}$}}
\def\capW{\mbox{\boldmath${\mathsf W}$}}

\def\i{\mbox{${\mathrm i}$}}
\def\mC{\mbox{\boldmath${\mathcal C}$}}
\def\mB{\mbox{${\mathcal B}$}}
\def\mE{\mbox{${\mathcal{E}}$}}
\def\mL{\mbox{${\mathcal{L}}$}}
\def\mK{\mbox{${\mathcal{K}}$}}
\def\mV{\mbox{${\mathcal{V}}$}}
\def\C{\mbox{\boldmath${\mathcal C}$}}
\def\E{\mbox{\boldmath${\mathcal E}$}}

\def\AAM{{\it Advances in Applied Mechanics }}
\def\ACME{{\it Arch. Comput. Meth. Engng.}}
\def\ARMA{{\it Arch. Rat. Mech. Analysis}}
\def\AMR{{\it Appl. Mech. Rev.}}
\def\ASCEEM{{\it ASCE J. Eng. Mech.}}
\def\ACTA{{\it Acta Mater.}}
\def\CMAME {{\it Comput. Meth. Appl. Mech. Engrg.}}
\def\CRAS{{\it C. R. Acad. Sci. Paris}}
\def\CRM{{\it Comptes Rendus M\'ecanique}}
\def\EFM{{\it Eng. Fracture Mechanics}}
\def\EJMA{{\it Eur.~J.~Mechanics-A/Solids}}
\def\IJES{{\it Int. J. Eng. Sci.}}
\def\IJF{{\it Int. J. Fracture}}
\def\IJMS{{\it Int. J. Mech. Sci.}}
\def\IJNAMG{{\it Int. J. Numer. Anal. Meth. Geomech.}}
\def\IJP{{\it Int. J. Plasticity}}
\def\IJSS{{\it Int. J. Solids Structures}}
\def\IngA{{\it Ing. Archiv}}
\def\JAM{{\it J. Appl. Mech.}}
\def\JAP{{\it J. Appl. Phys.}}
\def\JAE{{\it J. Aerospace Eng.}}
\def\JE{{\it J. Elasticity}}
\def\JM{{\it J. de M\'ecanique}}
\def\JMPS{{\it J. Mech. Phys. Solids}}
\def\JSV{{\it J. Sound and Vibration}}
\def\MACRO{{\it Macromolecules}}
\def\MMT{{\it Mech. Mach. Th.}}
\def\MOM{{\it Mech. Materials}}
\def\MMS{{\it Math. Mech. Solids}}
\def\MMT{{\it Metall. Mater. Trans. A}}
\def\MPCPS{{\it Math. Proc. Camb. Phil. Soc.}}
\def\MSE{{\it Mater. Sci. Eng.}}
\def\NATURE{{\it Nature}}
\def\NATUREM{{\it Nature Mater.}}
\def\PHIL{{\it Phil. Trans. R. Soc.}}
\def\PMPS{{\it Proc. Math. Phys. Soc.}}
\def\PNAS{{\it Proc. Nat. Acad. Sci.}}
\def\PRE{{\it Phys. Rev. E}}
\def\PRL{{\it Phys. Rev. Letters}}
\def\PRSL{{\it Proc. R. Soc.}}
\def\RIIT{{\it Rozprawy Inzynierskie - Engineering Transactions}}
\def\ROCK{{\it Rock Mech. and Rock Eng.}}
\def\QAM{{\it Quart. Appl. Math.}}
\def\QJMAM{{\it Quart. J. Mech. Appl. Math.}}
\def\SCIENCE{{\it Science}}
\def\SCRMAT{{\it Scripta Mater.}}
\def\SM{{\it Scripta Metall.}}
\def\ZAMM{{\it Z. Angew. Math. Mech.}}
\def\ZAMP{{\it Z. Angew. Math. Phys.}}
\def\ZVDI{{\it Z. Verein. Deut. Ing.}}

\def\salto#1#2{
[\mbox{\hspace{-#1em}}[#2]\mbox{\hspace{-#1em}}]}

\renewcommand\Affilfont{\itshape}
\setlength{\affilsep}{1em}
\renewcommand\Authsep{, }
\renewcommand\Authand{ and }
\renewcommand\Authands{ and }
\setcounter{Maxaffil}{2}

\newcommand*\diff{\mathop{}\!\mathrm{d}}

\newcommand\ringring[1]{%
  {
   \mathop{\kern0pt #1}\limits^{
     \vbox to-1.85ex{
       \kern-2ex 
       \hbox to 0pt{\hss\normalfont\kern.1em \r{}\kern-.45em \r{}\hss}%
       \vss 
     }
   }
  }
}